\documentclass{article}
\usepackage{graphicx} % Required for inserting images
\usepackage{amsmath}
\usepackage{url}
\usepackage{natbib}
\setcitestyle{authoryear,open={(},close={)}} 

\usepackage[switch, modulo]{lineno}
\usepackage{geometry}
 \usepackage{authblk}
\title{Groundwater feedbacks on ice sheets and subglacial hydrology}

\author[1]{Gabriel J. Cairns}
\author[2]{Graham P. Benham}
\author[1]{Ian J. Hewitt}

\affil[1]{Mathematical Institute, University of Oxford, Woodstock Road, OX2 6GG, Oxford, UK}
\affil[2]{School of Mathematics and Statistics, University College Dublin, Belfield, Dublin, Ireland}
\begin{document}
\maketitle

\begin{abstract}
The dynamics of many of Antarctica's glaciers are modulated by a hydrological system at the base of the ice. 
Sedimentary basins beneath the ice bed contribute to the water budget in this hydrological system by discharging or taking up water. 
However, sedimentary basins are not included in most current models of ice dynamics, and little is known about their effect. 
In this paper we develop an idealised model of a glacier whose sliding is coupled to a subglacial hydrological system that includes a sedimentary basin. 
We find that groundwater discharge (exfiltration) and recharge (infiltration) are controlled by the shape of the ice sheet and of the sedimentary basin, and that exfiltration promotes sliding whereas infiltration hinders it. 
Overall, the presence of a sedimentary basin leads to thicker and slower-flowing ice in the steady state. 
We also find that, when the ice sheet is undergoing retreat, groundwater infiltration near the grounding line can produce a feedback which decelerates this retreat.
%We also find that, when the ice sheet is undergoing retreat, groundwater exfiltration can lead to a positive feedback which accelerates this retreat. 
Our results shed light on the potential role and importance of Antarctic sedimentary basins, and how these might be incorporated into existing models of ice and subglacial hydrology.
\end{abstract}

\section{Introduction}
% Introduction and lit review
% Importance of Antarctic ice streams. Subglacial hydrology modelling. Previous approaches to sedimentary basins. 

The Antarctic ice sheet is the largest body of ice on the planet, with the potential to contribute an estimated 30 cm of sea-level rise by 2100 and up to 6.9 m by 2300 under high-emission climate scenarios \citep{seroussi2024evolution}. 
One of the key sources of uncertainty in this estimate is our understanding of the Antarctic subglacial hydrological system \citep{dow2022role, fricker2025antarctica}. 
This hydrological system originates mainly from water formed by melting at the base of the ice \citep{mccormack2022fine} and modulates the dynamics of many of Antarctica's ice streams by affecting how rapidly the ice slides over the bed \citep{engelhardt1990physical, morlighem2013inversion}. 

One potential important contributor to the subglacial hydrological system is the exchange of groundwater with sedimentary basins. 
Sedimentary basins underlie large portions of Antarctica, and host substantial volumes of groundwater \citep{gustafson2022dynamic, tankersley2022basement, li2022sedimentary, aitken2023antarctic, li2024crustal}. 
This groundwater is discharged into the hydrological system at the base of the ice in a process known as exfiltration (or infiltration, when water instead enters the sedimentary basin), which may contribute significantly to water budgets at the base of ice streams 
\citep{christoffersen2014significant}. 
However, little is currently known about the effect of these sedimentary basins on the subglacial hydrological system and hence the sliding of the ice, and the implications for contemporary ice sheet retreat.

Due to the difficulty of directly observing subglacial processes, mathematical modelling has emerged as a valuable tool in the study of subglacial hydrology \citep{dow2022role}, although the majority of current subglacial hydrology models do not include exfiltration from, or infiltration into, sedimentary basins. 
Various approaches consider water transport through subglacial conduits, a linked cavity network containing a water film (a ``cavity sheet"), or flow through porous sediments \citep{flowers2015modelling}. 
The most frequently used state-of-the-art models combine multiple components, for example by including both a cavity sheet and a conduit system  \citep[\textit{e.g.}][]{werder2013modeling, sommers2018shakti}. 
Such multi-component hydrology models have been used to model Antarctic subglacial hydrology for both individual glacier catchments \citep{dow2022antarctic, wearing2024characterizing, zhang2024role, hayden2025past} and continent-wide settings \citep{ehrenfeucht2025antarctic}. 

While some studies also include feedback of the ice on the hydrology \citep{cook2020coupled, pelle2024subglacial}, two-way coupled modelling of ice and hydrology remains infrequent in state-of-the art models due to its high computational demand and the large parameter space required \citep{dow2022role}. 
As a result, relatively idealised models are more popular as a means of gaining insight into the coupling between ice sheets and subglacial hydrology \citep[\textit{e.g.} ][]{lu2023coupling, kazmierczak2024fast, haseloff2025subglacial}. 
Such models may, for instance, work in two rather than three spatial dimensions, or consider only one component in a subglacial hydrology model (\textit{e.g.} a conduit or a cavity sheet). 

Some previous models of sedimentary basins have estimated exfiltration and infiltration driven by sediment compaction under idealised ice sheet retreat \citep{gooch2016potential, li2022sedimentary} and for contemporary ice sheet thinning rates \citep{robel2023contemporary}. 
Separately, a model of two-dimensional flow neglecting compaction has been used to estimate exfiltration and infiltration driven by spatial variations in horizontal groundwater flux \citep{cairns2025groundwater}. 
Although these studies find that exfiltration from both compaction and horizontal flux variation could contribute significantly to basal water budgets, they do not model the effect of this flux on the subglacial hydrological system and its coupling with the ice sheet. 

In this paper, we therefore seek to develop a simple mathematical model for ice sheet flow in one horizontal dimension, coupled to a subglacial hydrological system, which includes exfiltration and infiltration from a sedimentary basin due to both compaction and horizontal flux variation. 
This model enables us to investigate how groundwater contributes towards or against the sliding of the ice, and where this effect is most important. 
We also investigate the consequences for the hydrology and ice sheet of compaction-driven exfiltration and infiltration during retreat. 

We introduce the governing equations of this model in Section 2, and non-dimensionalise to find that the influence of the groundwater flow in the sedimentary basin is characterised by the parameters $K$ and $\Sigma$, effectively a dimensionless permeability and specific storage. 
Section 3 describes the steady states of the model when the feedback from groundwater flow on the hydrological system (and therefore the ice sheet) is weak. 
In Section 4, we then investigate the effect of strong groundwater flow coupling on these steady states, finding that the presence of a sedimentary basin can lead to thicker and slower-flowing ice in the steady state.
In Section 5, we consider transient solutions when the ice is retreating or advancing to a new steady state, and show in particular that compaction-driven infiltration and exfiltration can decelerate ice sheet retreat and advance respectively.  
In Section 6, we conclude with a discussion of our results and possible developments of the model. 

\section{Model development}
%Fig: schematic. Ice sheet model. Hydrology model. Sedimentary basin model (derivation consigned to appendix). 

\begin{figure}
    \centering
    \includegraphics[width=0.85\linewidth]{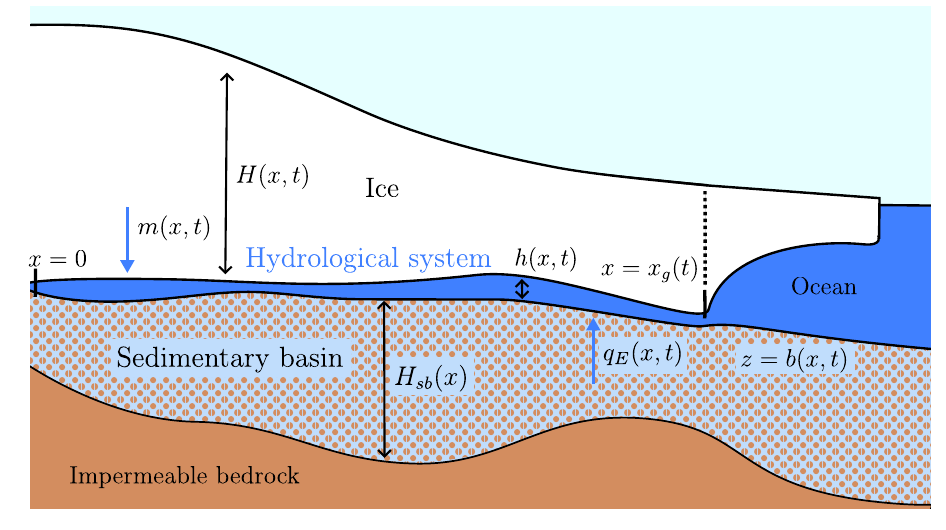}
    \caption{Schematic of the model, featuring an ice sheet with thickness $H(x,t)$ and hydrological system with effective thickness $h(x,t)$, above the bed elevation $b(x,t)$, with a sedimentary basin of thickness $H_{sb}(x)$ beneath. The grounded portion of the ice occupies $0<x<x_g(t)$. Melting of the ice supplies a flux of water $m(x,t)$ to the hydrological system, and exfiltration (or infiltration if $q_E<0$) from the sedimentary basin provides a  flux $q_E(x,t)$.}
    \label{fig:schematic}
\end{figure}

We consider a model with one horizontal dimension in addition to the vertical, in which the ice flows along the $x$-direction, as shown in Figure \ref{fig:schematic}. 
The ice has thickness $H(x,t)$, and lies on a bed at $z=b(x,t)$, where $z=0$ refers to sea level. 
The ice divide is located at $x=0$, and we assume this point also corresponds to the onset of the subglacial hydrological system. 
% * do we need to justify
The grounded ice flow transitions into a floating ice shelf at the grounding line $x=x_g(t)$, a point which must be determined as part of the solution. 

Beneath the ice is a subglacial hydrological system. 
For simplicity, we model this system using the `effective' water thickness $h(x,t)$, which represents the average volume of water per unit area. 
This follows the approach of treating the subglacial hydrological system as a macroporous cavity sheet  \citep[\textit{e.g.}][]{alley1996towards, creyts2009drainage, lebrocq2009subglacial}, which achieves comparable results to more complex models over long timescales \citep{defleurian2018shmip}. 
By a `macroporous cavity sheet', we mean that the water resides in interconnected cavities between protrusions in the bed (\textit{i.e.} the sedimentary basin rock), which directly connect with the ice. (These cavities are however much larger than the pores in the sedimentary rock itself, hence the term `macroporous'.) 
The effective water thickness $h(x,t)$ is a larger-scale quantity, presumed to be averaged over a lengthscale much larger than the individual cavities. 
Melting of the ice supplies a flux $m(x,t)$ of water to the hydrological system. 

Our model also includes a sedimentary basin, with thickness $H_{sb}(x)$. 
We assume that the effective water thickness $h(x,t)$ is much smaller than $H(x,t)$ and $H_{sb}(x)$, so that the top of the sedimentary basin is at approximately the same position $z=b(x,t)$ as the base of the ice sheet. 
We use $q_E$ to denote the exfiltration (or infiltration when $q_E<0$ ), the flux of water from the sedimentary basin into the hydrological system or vice versa. 

\subsection{Ice sheet}

We adopt the ``shallow-shelf approximation" for ice flow, a standard model used to model relatively fast-flowing ice streams, where the ice moves primarily by sliding over the bed rather than through internal shearing \citep{muszynski1987coupled, macayeal1989large}. 
This model assumes the ice has vertically-uniform velocity $u(x,t)$, so that the ice flux is given by $q(x,t) = Hu$. 
%This assumption is appropriate when the ice moves primarily by sliding over the bed rather than through internal shearing, and thus is appropriate for our investigation, in which subglacial hydrology affects the ice through basal sliding. 
Conservation of mass for the ice sheet therefore gives
\begin{equation}\label{eq:massi}
  \frac{\partial H}{\partial t} +   \frac{\partial}{\partial x}{} \left(H u\right) = a,
\end{equation}
where $a(x,t)$ is the net rate of ice mass accumulation by compaction of snow. 
For simplicity, we assume $a$ is a known positive constant. 

The ice velocity $u$ is determined by momentum conservation, which is given by
\begin{equation}\label{eq:momi}
  \frac{\partial}{\partial x}{} \left(2 A^{-1/n} H \left|  \frac{\partial u }{\partial x} \right|^{1/n-1}\frac{\partial u }{\partial x} \right) - \tau_b(u,N)  - \rho_i g H   \frac{\partial}{\partial x}{}\left(H+b\right) = 0.
\end{equation}
Here the first term is the divergence of the depth-integrated extensional stress in the ice. 
This stress is determined by the nonlinear rheology of ice, known as Glen's law. 
The constant $A$ is the (vertically-averaged) Glen's law coefficient, which is well-constrained by measurements, albeit temperature-dependent. Our chosen value of $A =1 \times 10^{-25}$ s$^{-1}$ Pa$^{-3}$ corresponds to an average ice temperature of roughly -25$^\mathrm{o}$ C. 
We take the corresponding exponent $n=3$, as is commonly assumed in ice sheet models \citep{cuffey2010physics}. 
%Since in reality the temperature and hence $A$ vary throughout the depth of the ice, the value of $A^{-1/n}$ in Equation \eqref{eq:momi} is strictly speaking a vertical average. 
% Check explanation

The central term $\tau_b(u,N)$ of Equation \eqref{eq:momi} is the basal shear stress, \textit{i.e.} the frictional force per unit area imposed on the ice by the bed. 
We assume this stress depends only on the ice velocity $u(x,t)$ and the `effective pressure' $N(x,t)$ within the subglacial hydrological system. 
The latter is defined as the difference between the overburden imposed by the ice and the water pressure $p_w$ in the subglacial water layer
\begin{equation}
N = \rho_i g H - p_w,
\end{equation}
where $\rho_i$ is the density of ice and $g$ is gravitational acceleration. 
Like $h$, $N$ is a macroscopic quantity over lengthscales larger than individual cavities. 
The effective pressure is zero when the water pressure is sufficient for the ice to float. %, and is positive when the water bears a fraction of the weight of the ice. 
The third term of Equation \eqref{eq:momi} is the driving stress, which corresponds to the force imposed on the ice by its own weight. 
% A fuller derivation of Equation \eqref{eq:momi} is provided in Appendix A. 

We model the basal shear stress using a `regularised Coulomb' sliding law \citep{schoof2005effect, gagliardini2007finite, zoet2020slip},
\begin{equation}\label{eq:rc}
\tau_b = C_C N \left(\frac{ |u| }{|u| + \left(C_C N / C_W\right)^{1/m_W} }\right)^{m_W} \text{sgn}(u),
\end{equation}
where $C_W$ and $C_C$ are empirical constants, and the exponent $m_W$ is usually taken to be related to the Glen's law exponent by $m_W =1/ n = 1 / 3$. 
The law \eqref{eq:rc} implies that the basal stress $\tau_b$ is an increasing function of $N$, with the limiting behaviour
\begin{equation}
\tau_b \sim \begin{cases}
C_W  {|u|}^{m_W} \text{sgn}(u) \quad  &C_C N \gg C_W u^{m_W},  \\
C_C N \text{sgn}(u) \quad  &C_C N \ll C_W u^{m_W}.
\end{cases}
\end{equation}
The use of the law \eqref{eq:rc} has been supported over a range of bed types by numerical modelling \citep{gagliardini2007finite, joughin2019regularized,  helanow2021slip} and empirical observations \citep{zoet2020slip}. 
For high effective pressures, the law \eqref{eq:rc} recovers the `Weertman' sliding law frequently used in ice sheet models where hydrology has little effect on the sliding of the ice \citep{weertman1957sliding, fowler1981theoretical}, and for low effective pressures the `Coulomb' sliding law is recovered, in which the friction is proportional to $N$. 

To solve the above model for $H(x,t)$, $u(x,t)$ and the free boundary $x_g(t)$, Equations \eqref{eq:massi} and \eqref{eq:momi} must be combined with spatial boundary conditions. 
The first of these is a prescription of no ice flux at the ice divide:
\begin{equation}\label{eq:x0i}
u =0 \quad \text{at} \quad x=0.
\end{equation}
The other two boundary conditions are imposed at the grounding line. 
The first of these is that the ice must achieve flotation, \textit{i.e.} the weight of the ice equals that of the displaced water
\begin{equation}\label{eq:Hixg}
\rho_i g H = - \rho_w g b \quad \text{at} \quad x=x_g,
\end{equation}
using $\rho_w$ to denote the density of water. 
The second is the stress condition
\begin{equation}\label{eq:stressxg}
2 A^{-1/n} H \left| \frac{\partial u}{\partial x}\right|^{1/n} \frac{\partial u}{\partial x} = \frac12 \rho_i \left(1 - \frac{\rho_i}{\rho_w}\right) g H^2  \quad \text{at} \quad x=x_g.
\end{equation}
A derivation of Equation \eqref{eq:stressxg}, which results from considering the physics of an unbuttressed floating ice shelf beyond the grounding line can be found in \citet{schoof2007ice}. 

\subsection{Subglacial hydrological system}
%Mass conservation equation. Explanation of terms. Constitutive equation. Boundary conditions.

The evolution of the effective water thickness $h(x,t)$ in the hydrological system is governed by an equation of mass conservation
\begin{equation}\label{eq:hpre}
  \frac{\partial h}{\partial t} +   \frac{\partial Q_h}{\partial x} = m + q_E.
\end{equation}
Here $Q_h$ is the flux of subglacial water. 
The source term $m$ describes the supply of water to the hydrological system due to melting of the ice. 
The exfiltration $q_E$ from the underlying sedimentary basin also appears as a source term. 

For simplicity, we treat $m$ as a constant parameter, whose size may be estimated by considering the melting produced by inferred geothermal heat fluxes \citep{mccormack2022fine}, and whose effect on the ice mass balance is included in the net accumulation $a$. %, and which does not contribute significantly to the overall ice mass balance. 
Assuming a latent heat of fusion  $L=3.3 \times 10^5$ J k$g^{-1}$ and a density $\rho_i=900$ kg m$^{-3}$, a melt rate of $m=1$ mm yr$^{-1}$ requires a net heat flux $Q=\rho_iLm$ of around 9 mW m$^{-2}$. Since estimates of geothermal heat flux beneath Antarctica range from 30--150 mW m$^{-2}$ \citep{burton2020geothermal, staal2021antarctic}, a melt rate of a few mm yr$^{-1}$ is reasonable allowing for some heat conduction into the ice. 
In our assumption of constant $m$, we have neglected the role of frictional heating in meltwater production, which in reality is likely to be significant, if not dominant, in the basal energy budget. 
In Section 6 we discuss the implications of neglecting frictional heating, and possible extensions of the model to include it.

For the flux $Q_h$, we prescribe a Darcy-like law % wherein the flux is proportional to the effective water thickness multiplied by the gradient of the hydraulic potential
\begin{equation}\label{eq:qh}
Q_h = - \frac{k}{\eta} h  \frac{\partial\psi}{\partial x} ,
\end{equation}
where the hydraulic potential $\psi$ is defined as the difference between the water pressure and hydrostatic pressure
\begin{equation}\label{eq:psi}
\psi = p_w + \rho_w g b = \rho_i g H - N + \rho_w g b.
\end{equation}
In Equation \eqref{eq:qh}, the constant $k$ is the effective permeability of the subglacial hydrological system and $\eta$ is the viscosity of water. 
The former is difficult to measure and in theory depends on properties of the macroporous cavity-sheet network such as its average cavity size and connectivity. 
%However, we can estimate a value of $k$ if we assume that the terms in Equation \eqref{eq:hpre} are of comparable size. 
However, we can estimate a value of $k$ based on the fact that the subglacial horizontal water flux must be large enough to evacuate the meltwater produced by geothermal heating. Our estimate of $k$ does not substantially affect the model, since decreasing $k$ causes the effective water thickness $h$ to increase, and vice versa, so that the combination $k h$ remains roughly consistent (see Appendix A).
%\citet{haseloff2025subglacial} consider a range of values equivalent to $k$ between $10^{-9}$ and $10^{-6}$ m$^2$, for which our estimate of $4 \times 10^{-8}$ m$^2$ (see Table 1 in Appendix) represents an intermediate value. 
For simplicity, we ignore the density difference between fresh basal meltwater and saline ocean water and take all subglacial water to have the same density $\rho_w$. 

%In order to close the subglacial hydrology problem,
Finally, we require a `closure relation' linking $h$ to $N$, which we assume can be written as a decreasing function
\begin{equation}\label{eq:Nhgen}
h = h(N).
\end{equation}
Equation \eqref{eq:Nhgen} is motivated by models of subglacial cavities, in which the rate of cavity closure by viscous creep increases with both effective pressure and the spacing between protruding subglacial obstacles. 
The latter increases with water film thickness $h$ as the obstacles become more submerged. 
Assuming that opening and closure of cavities are in quasi-equilibrium and approximating the opening rate as constant leads to a decreasing relationship of the form \eqref{eq:Nhgen} 
\citep[\textit{e.g.}][]{creyts2009drainage, schoof2012flotation, hewitt2013seasonal, werder2013modeling, kyrke2014subglacial}. 
A relationship of the form \eqref{eq:Nhgen} also arises in models of subglacial till, if the void ratio of the till is taken as a decreasing function of the effective pressure \citep{tulaczyk2000basal, haseloff2025subglacial}.
% * effective pressure / effective stress equivalence

% This relation physically expresses how the water pressure in the cavities resists the closure of these cavities due to viscous creep of the ice. 
% The function is decreasing because we expect cavity closure to drive water out when the water pressure $p_w$ is low, \textit{i.e.} when $N$ is high. 
For the examples in this paper, we use the idealised relationship
\begin{equation}\label{eq:Nh}
%N = \frac{E}{h},
h = \frac{E}{N},
\end{equation}
where $E$ is a constant representing the `stiffness' of the cavity sheet system. 
% Owing to the idealised nature of this model, we must select an appropriate value for $E$ based on realistic values for $N$ and $h$. 
A particular feature of Equation \eqref{eq:Nh} is that the effective water thickness $h \to \infty$ as $N \to 0$. 
This means that as the ice approaches flotation, an unbounded amount of water can be stored beneath. 
An idealisation of the form \eqref{eq:Nh} is reasonable when effective water thicknesses are much smaller than the height of the largest bed obstacles. 
A more physically accurate choice would have $N=0$ for $h \geq h_c$, where $h_c$ is the maximum height of bed obstacles, since the ice should achieve flotation once all obstacles are submerged. 
However, we assume $h_c$ can be treated as infinite, since $h$ remains finite away from $x_g$ in the above solutions. % large values of $h$ are limited to a small region close to the grounding line. 

%This means that when the ice is at flotation, an unbounded amount of water can be stored beneath. 

Equation \eqref{eq:hpre} requires two spatial boundary conditions. 
As the first of these, we impose no flux of groundwater at the ice divide, where we assume the hydrological system begins,
\begin{equation}\label{eq:hx0}
Q_h = \frac{k}{\eta} h   \frac{\partial \psi}{\partial x}  = 0 \quad \text{at} \quad x=0.
\end{equation}
For the second, we require that the effective pressure is zero at the grounding line
\begin{equation}\label{eq:Nxg}
N = 0 \quad \text{at} \quad x=x_g,
\end{equation}
which is a consequence of the flotation condition \eqref{eq:Hixg}. 
%This condition comes from imposing continuity of the water pressure across the grounding line, using the fact that the overlying ice is at flotation, as stipulated in Equation \eqref{eq:Hixg}. 
% Combining Equation \eqref{eq:Nxg} with Equation \eqref{eq:Nh} implies that $h \to \infty$ as $x \to x_g$, which reflects the fact that the subglacial cavity network becomes continuous with the much deeper ocean. 

\subsection{Sedimentary basin}
%Governing equation. Eliminating $q_E$. %Including a potential compaction term?

We determine the exfiltration $q_E$ by modelling the groundwater flow in the underlying sedimentary basin. 
The basin is assumed to have thickness $H_{sb}(x)$, so that it occupies the region $b-H_{sb}<z<b$. 
%We begin with the groundwater flow equation including one-dimensional vertical compaction (\textit{e.g.} \citet{lemieux2008dynamics, li2022sedimentary} (in supplement); \citet{, robel2023contemporary}), using $p_{sb}(x,z,t)$ to denote the water pressure in the sedimentary basin,
We begin with the groundwater flow equation for uniaxial strain \citep[\textit{e.g.}][]{lemieux2008dynamics, robel2023contemporary}, using $p_{sb}(x,z,t)$ to denote the water pressure in the sedimentary basin,
\begin{equation}\label{eq:groundwater}
\frac{S_s}{\rho_w g}   \frac{\partial {p_{sb}}}{\partial t} =   \frac{\partial}{\partial x} \left(\frac{k_{sb}}{\eta}   \frac{\partial {p_{sb}}}{\partial x}\right) +   \frac{\partial}{\partial z} \left(\frac{k_{sb}}{\eta} \left(  \frac{\partial p_{sb}}{\partial z}+\rho_w g\right)\right) + \frac{S_s \xi \rho_i}{\rho_w}  \frac{\partial H}{\partial t} .
\end{equation}
Here $S_s$ is the one-dimensional specific storage, $k_{sb}$ is the permeability of the sedimentary basin (assumed uniform and isotropic), and $0<\xi \leq 1$ is the one-dimensional loading efficiency. 
%The final term depends on the water pressure $p_w$ in the subglacial hydrological system, \textit{i.e.} at the top of the sedimentary basin. 
%The final term depends on the water pressure $p_w = \rho_igH -N$ in the subglacial hydrological system, which determines the overall loading stress on the sedimentary basin. 
The final term is the rate of change of the total vertical loading stress on the sedimentary basin, which is given by the overburden pressure $\rho_i g H$. %This is because we assume that the ice sheet is supported by protrusions in the bed between water-filled cavities. 
% However, w
We assume that the water in the sedimentary basin is continuous with the hydrological system, so that $p_{sb}=p_w$ at $z=b$. 
(Since this formulation generally allows for a positive effective pressure $N=\rho_i g H - p_w$, it also assumes that premelting occurs between the ice and the bed, \textit{i.e.} small-scale thermodynamic effects ensure that the force balance is satisfied in a nanofilm of water between the ice and bed \citep{gerber2022stress}.) 

Equation \eqref{eq:groundwater} follows from water mass conservation, assuming the water and porous medium are weakly compressible, combined with Darcy's law for porous flow. 
This equation also assumes the strain is uniaxial, which is valid provided the overburden varies over a lengthscale larger than the thickness of the sedimentary basin. 
A full derivation can be found in \citet{lemieux2006impact}. 

In using Equation \eqref{eq:groundwater}, we have regarded the sedimentary basin as a lithified rock, which undergoes small deformations that can be modelled using poro-elastic theory. 
The resulting equation therefore differs from previous models for the compaction of subglacial till, which allow large deformations and non-linear strains \citep{tulaczyk2000basal, tulaczyk2000basal2, tulaczyk2001estimates}. 
While such deformations may be possible in the unlithified upper layer of sediment, we assume that this layer is negligibly small compared to the sedimentary basin as a whole.

%for weakly comp, including the effect of compaction of the porous medium and compressibility of the water itself (which is elsewhere negligible) via linear poroelasticity theory. 

% Equation \eqref{eq:groundwater} is derived from water mass conservation, including the effect of compaction of the porous medium and compressibility of the water itself (which is elsewhere negligible) via linear poroelasticity theory. 
% This is combined with Darcy's law for flow in porous media. 
% A derivation of this equation from first principles can be found in \citet{lemieux2006impact}. 
% * Explain more about these constants?

The exfiltration (or infiltration) is given by the normal Darcy flux at the top of the sedimentary basin
\begin{equation}\label{eq:qE0}
q_E = -\left. \frac{k_{sb}}{\eta}\left(  \frac{\partial {p_{sb}}}{\partial z}+\rho_w g - \frac{\mathrm{d}b}{\mathrm{d}x}   \frac{\partial {p_{sb}}}{\partial x} \right) \right|_{z=b(x)}. 
\end{equation}
To determine this, we impose a boundary condition of no normal Darcy flux through the bottom of the sedimentary basin, where a layer of impermeable bedrock is assumed, such that
\begin{equation}\label{eq:bottomflux}
0 = -\left. \frac{k_{sb}}{\eta}\left(  \frac{\partial{p_{sb}}}{\partial z}+\rho_w g - \left(\frac{\mathrm{d}b}{\mathrm{d}x} -   \frac{\mathrm{d}{H_{sb}}}{\mathrm{d} x}\right)   \frac{\partial{p_{sb}}}{\partial x} \right) \right|_{z=b(x)-H_{sb}(x)}. 
\end{equation}
We may then integrate Equation \eqref{eq:groundwater} over the depth of sedimentary basin to obtain
\begin{equation}\label{eq:groundwaterint}
q_E =   \frac{\partial}{\partial x}{} \left[ \frac{k_{sb} }{\eta}  \int_{b-H_{sb}}^b   \frac{\partial {p_{sb}}}{\partial x} \, \mathrm{d}z \right] + \int_{b-H_{sb}}^b  \left(\frac{S_s \xi \rho_i}{\rho_w}    \frac{\partial H}{\partial t} - \frac{S_s}{\rho_w g}   \frac{\partial {p_{sb}}}{\partial t} \right)\, \mathrm{d}z  ,
\end{equation}
where we have used Leibniz's integral rule.

We now make the assumption that the sedimentary basin is much longer than it is deep. 
This assumption is frequent in groundwater flow problems, known as the Dupuit approximation \citep{bear2013dynamics}, and is consistent with inferred Antarctic sedimentary basin geometry \citep{tankersley2022basement, li2024crustal}. 
(As well as requiring a small aspect ratio, the Dupuit assumption requires that the quantity $S_s \eta [H]^2 / k_{sb} \rho_w g[t]$ is small, where $[H]$ and $[t]$ are scales for sedimentary basin thickness and time. This is true on the long timescales of our consideration provided that $k_{sb}$ is on the larger side of our range of values. The case where this assumption fails is discussed in Appendix B.)

This approximation implies that the water pressure is roughly hydrostatic %to leading order in the aspect ratio 
\begin{equation}\label{eq:hydrostatic}
p_{sb} \approx p_w + \rho_w g (b - z).
\end{equation}
Substituting of \eqref{eq:hydrostatic} into \eqref{eq:groundwaterint} gives
%\begin{equation}
% q_E =    \frac{\partial}{\partial x}{} \left[ \frac{k_{sb}}{\eta} H_{sb} \left(   \frac{\partial}{\partial x}{p_w} + \rho_w g \frac{\mathrm{d}b}{\mathrm{d}x}  \right) \right] - \frac{S_s (1-\xi)}{\rho_w g} H_{sb}   \frac{\partial}{\partial t}{p_w}.
% \end{equation}
%Finally, we can substitute $p_w = \rho_i g H - N$ to obtain an equation for the exfiltration in terms of these variables
%Finally, substituting 
%and letting 
%$p_w = \rho_i g H - N$ gives
\begin{equation}\label{eq:qE}
q_E =    \frac{\partial}{\partial x}{} \left[ \frac{k_{sb}}{\eta} H_{sb}   \frac{\partial}{\partial x}{} \left( \rho_i g H - N + \rho_w g b \right) \right] - \frac{S_s}{\rho_w g} H_{sb}  \left( (1- \xi) \rho_i g   \frac{\partial H}{\partial t}  - \frac{\partial  N}{\partial t} \right).
%\frac{S_s (1- \xi) \rho_i}{\rho_w } H_{sb}   \frac{\partial H}{\partial t}{}.
\end{equation} 

The exfiltration in Equation \eqref{eq:qE} consists of a `topographic' and `compaction' term respectively. 
The topographic term is (minus) the divergence of the depth-integrated horizontal Darcy flux through the aquifer, which leads to exfiltration when this total flux is decreasing in the direction of flow. 
%This term is positive when the hydraulic potential gradient is becoming less steeply negative (\textit{e.g.} due to the ice sheet profile flattening) or the sedimentary basin thickness is becoming shallower in the direction of flow. 
The size of this term is controlled by the sedimentary basin permeability $k_{sb}$. 
Although $k_{sb}$, like $k$, is difficult to measure directly in Antarctica, it can be estimated based on values from other similar sedimentary aquifers. Values of $k_{sb}$ for coarse-grained sedimentary rocks range from $10^{-17}$--$10^{-12}$ m$^2$, and $10^{-19}$--$10^{-13}$ m$^2$ for glacial tills \citep{freeze1979groundwater}.  

% \citep{gooch2016potential, li2022sedimentary, robel2023contemporary}, estimated exfiltration fluxes \citep{christoffersen2014significant}, or models of other processes such as seawater intrusion \citep{cairns2025groundwater}. 

The size of the compaction term is controlled by the specific storage $S_s$, and also depends on the loading efficiency $\xi$. 
To understand the origin of this term, we may write it as
\begin{equation}\label{eq:qEc}
q_{Ec} = 
\frac{S_s}{\rho_w g} H_{sb}  \left( \xi \frac{\partial  N}{\partial t} -(1- \xi)  \left( \rho_i g   \frac{\partial H}{\partial t} -\frac{\partial  N}{\partial t}  \right) \right).
\end{equation}
The term with prefactor $\xi$ comes from the compression or decompression of the porous medium, and depends on the effective stress in the porous medium at the bed, which in this case equals the effective pressure. %(a consequence of premelting, as mentioned above). 
This term produces exfiltration when the effective pressure increases ($\partial N / \partial t>0$), as the porous medium contracts, expelling water. 
The term with prefactor $(1-\xi)$ is due to weak compressibility of the water. 
This term produces exfiltration when the water pressure decreases, as the water then expands, causing some water to be discharged. 
Generally, we expect that the effective pressure is small compared to $(1-\xi)$ times the overburden, and so the latter dominates, giving
\begin{equation}
q_{Ec} \approx \frac{\rho_i S_s }{\rho_w} H_{sb} (1- \xi)    \frac{\partial H}{\partial t}.
\end{equation}
This leads to exfiltration when the ice sheet is thinning, and infiltration when the ice sheet is thickening \citep{robel2023contemporary}. 
For a strongly deformable porous medium such as unconsolidated till, the loading efficiency $\xi$ would be close to 1, and the `effective stress' term would dominate instead. 
As discussed above, however, we regard the sedimentary basin as a lithified rock, for which it is reasonable to assume $\xi <1$.
%The size of the compaction term is controlled by the specific storage $S_s$, and secondarily the loading efficiency $\xi$. 
 
% The compaction term accounts for deformation of the porous medium, and is positive when the loading stress $H$ is decreasing. %, due to the fact that the weakly compressible pore water expands more during unloading than the porous medium dilates. 
% %During unloading, the weakly compressible pore water expands at a rate 
% This is because unloading allows the porous medium to dilate, reducing the water pressure. 
% The weakly compressible water expands in response to this pressure change, leading to exfiltration. 
% The size of the compaction term is controlled by the specific storage $S_s$, and secondarily the loading efficiency $\xi$. 

% The loading efficiency $\xi$ represents the compressibility of the porous medium compared to the total compressibility of the porous medium and the  water. 
% For a highly deformable medium, such as unlithified sediment, the compressibility of the medium is much larger than water, so that $\xi \approx 1$, in which case the compaction-driven exfiltration vanishes. 
% As discussed above, however, we regard the sedimentary basin as a lithified rock, for which it is reasonable to assume $\xi <1$.

Like the permeability $k_{sb}$, $S_s$ and $\xi$ are hard to measure directly, but can be estimated based on the properties of other similar sedimentary basins. For formerly glaciated aquifers in North America, \citet{neuzil2014ice} measure specific storages ranging from $1 \times 10^{-7}$ to $3 \times 10^{-5}$ m$^{-1}$, and loading efficiencies between $0.3$ and $1.0$, though generally around 0.75--0.95. 
However, \citet{stone1997inversion} use a drainage model with borehole data to infer a specific storage of $ 1\times 10^{-4}$ m$^{-1}$ for subglacial till. 
A reasonable range is therefore $S_s = 1\times 10^{-7}$--$1\times 10^{-4}$ m$^{-1}$, although in Section 5 we use exaggerated values up to $S_s = 6.8\times 10^{-4}$ m$^{-1}$ for illustrative purposes. We use a value $\xi=0.8$ (see Appendix A, Table 1).

% and on realistic scales for $q_E$ \citep{robel2023contemporary}. 

% Both $k_{sb}$ and $S_s$ (as well as $\xi$) are difficult to measure directly. 
% However, their size can be estimated based on the properties of similar sedimentary aquifers, as discussed in the following subsection. 

The formulation of groundwater flow and exfiltration here neglects certain thermodynamic effects, which fall beyond our current scope. 
For instance, at high effective pressures (larger than $10^4$ Pa), ice could infiltrate the porous bed to form a frozen fringe \citep{rempel2008theory}. 
This reduces the permeability of the frozen region significantly, but also drives a groundwater flow in order to satisfy the required rate of freeze-on. 
Moreover, the frozen region may inhibit sliding of the ice \citep{meyer2018freeze, hansen2024presence}. 
We discuss the implications of thermodynamic effects further in Section 6.

% Other drivers of exfiltration are possible, but fall beyond our current scope. 
% For example, basal freeze-on of subglacial water can drive exfiltration, but requires consideration of the energy balance at the ice bed \citep{christoffersen2003thermodynamics}. 

% Other than topographic and compaction effects, exfiltration can be driven by other phenomena, such as basal freeze-on of subglacial water \citep{christoffersen2003thermodynamics}. 
% However, modelling of such exfiltration again requires consideration of the energy balance at the base of the ice, which falls beyond our current scope. 

% Substituting Equations \eqref{eq:qh}, \eqref{eq:Nhgen} and \eqref{eq:qE} into Equation \eqref{eq:hpre} gives the combined hydrology equation
% \begin{equation}\label{eq:hqE2}
%   % \frac{\partial {h}}{\partial t} =   \frac{\partial}{\partial x}{} \left[ \frac{k h + k_{sb} H_{sb}}{\eta}   \frac{\partial}{\partial x}{} \left( \rho_i g H - N + \rho_w g b\right)  \right] \\ - \frac{S_s (1- \xi) \rho_i}{\rho_w } H_{sb}  \frac{\partial H}{\partial t}{} + m.
%  \frac{\partial {h}}{\partial t} =   \frac{\partial}{\partial x}{} \left[ \frac{k h + k_{sb} H_{sb}}{\eta}   \frac{\partial}{\partial x}{} \left( \rho_i g H - N + \rho_w g b\right)  \right] \\ - \frac{S_s}{\rho_w g} H_{sb}  \left( (1- \xi) \rho_i g   \frac{\partial H}{\partial t}  - \frac{\partial  N}{\partial t} \right) + m.
% \end{equation}
% * Overlong equation

\subsection{Non-dimensionalisation} 
We can gain insight into the model of ice and subglacial hydrology  described above by non-dimensionalising the governing equations and boundary conditions. 
That is, for each dimensional variable, we set
\begin{equation}
H = [H]\hat{H}, \quad x = [x]\hat{x}, \quad \dots
\end{equation}
where square brackets $[\cdot]$ denote the constant dimensional scales, and hats denote dimensionless variables. 
We then drop all hats from the notation. 
From here onwards, we work in dimensionless variables unless stated otherwise. 
However, we continue to use dimensional variables in figures and captions. 
The full process of non-dimensionalisation, including our choice of scales and the values used, is detailed in Appendix A. 

After non-dimensionalising, Equations \eqref{eq:massi} and \eqref{eq:momi} in the ice sheet model become
\begin{equation}\label{eq:massi*}
  \frac{\partial H}{\partial t} +   \frac{\partial}{\partial x}{} (H u) = a,
\end{equation}
\begin{equation}\label{eq:momih*}
4 \varepsilon   \frac{\partial}{\partial x}{} \left( H \left|  \frac{\partial{u}}{\partial x}\right|^{1/n-1} \frac{\partial{u}}{\partial x} \right) - \mathcal{C} N \left(\frac{ |u| }{|u| + (\mathcal{C} N )^{1/m_W} }\right)^{m_W} \text{sgn}(u)  - H   \frac{\partial}{\partial x}{}(H+b) = 0.
\end{equation}
The dimensionless parameter $\varepsilon \ll 1 $ represents the size of the extensional stress term compared to the basal stress. 
Although the extensional stress term is generally small, it becomes important near the grounding line, due to the stress balance boundary condition \eqref{eq:stressxg}. 
The result is that solutions of this ice-sheet model usually feature an `extensional stress boundary layer', a small region near the grounding line in which the extensional stress terms become important in Equation \eqref{eq:momih*}, while these terms remain negligible elsewhere \citep{schoof2007ice, tsai2015marine}. 
The dimensionless friction coefficient $\mathcal{C}$ is $O(1)$, and determines how strongly the sliding of the ice is coupled to the subglacial hydrology.

In considering the exfiltration, we non-dimensionalise $H_{sb}$ using the ice sheet thickness scale $[H]$, since sedimentary basins, like the ice sheet, are typically hundreds to thousands of metres deep \citep{gustafson2022dynamic, tankersley2022basement}. 
The non-dimensional version of the subglacial hydrology equation \eqref{eq:hpre} is then given by
\begin{multline}\label{eq:hyd*2}
\lambda   \frac{\partial h}{\partial t} =   \frac{\partial}{\partial x}{} \left[ \left(h + K H_{sb}\right)    \frac{\partial}{\partial x}{} \left( (1-\delta)  H - \mathcal{E} N +  b\right)  \right] + m \\ - \Sigma H_{sb}   \left( (1-\delta)(1-\xi)\frac{\partial H}{\partial t}  - \mathcal{E}\frac{\partial N}{\partial t}\right).
\end{multline}
For a given ice thickness $H$, Equation \eqref{eq:hyd*2} is a nonlinear diffusion equation for $N$, containing several dimensionless parameters. 
The parameter $\delta = 1- \rho_i / \rho_w \approx 0.1$ accounts for the density difference between water and ice, and is known. 
The dimensionless timescale $\lambda \ll 1 $ represents the ratio of the subglacial hydrology's natural timescale to that of the ice sheet. 
The fact that $\lambda$ is small suggests the subglacial hydrological system generally evolves much faster than the ice sheet in the absence of compaction-driven exfiltration, but the presence of $\partial N / \partial t$ in the compaction term produces a slower evolution on a timescale $\Sigma \mathcal{E}$. 

The parameter $\mathcal{E}$ represents the characteristic size of the effective pressure relative to that of the ice overburden. 
Since borehole observations by \citet{engelhardt1997basal} in West Antarctica found effective pressures to be a small percentage of the overburden ($<0.2$ MPa beneath 1000 metres of ice, an overburden of around 9 MPa), we expect $\mathcal{E}$ to be small. In particular, we take a value $\mathcal{E}=0.05$. Since varying $\mathcal{E}$ is equivalent to varying other parameters of the model (see Appendix A), it suffices for our analysis to consider this single value. 

% The parameter $\mathcal{E}$ represents the characteristic size of the effective pressure relative to that of the ice overburden, which we expect to be fairly small based on existing models of effective pressure beneath Antarctica \citep[\textit{e.g.}][]{ehrenfeucht2025antarctic}. 
% We take $\mathcal{E}=0.05$, based on the the previous modelling assumption by \citet{bueler2009shallow} that subglacial effective pressure is at most 5\% of the overburden. 
The permeability ratio $K$ 
\begin{equation}
%K = \frac{k_{sb} [H]}{k [h]} = \frac1{ [m][x] } \left([H] \frac{k_{sb} \rho_i g}{\eta} \frac{[H] }{[x]} \right),
K = \frac{k_{sb} [H]}{k [h]} = \frac1{ [m][x] } \left([H] \frac{k_{sb} \rho_w g}{\eta} \frac{[H] }{[x]} \right)
\end{equation}
is the ratio of a typical horizontal flux through the sedimentary basin to that of the subglacial cavity sheet. 
The size of $K$ determines the flux of water transported through the sedimentary basin rather than the cavity sheet, along with the overall scale of topographic exfiltration. 
Using a range $k_{sb}=1 \times 10^{-17}$--$1 \times 10^{-12}$ m$^2$ gives $K$ ranging from $6.1 \times 10^{-5}$ to $6$. 
In this paper, we will generally focus on higher values from this range, as the impact of horizontal groundwater flow is small otherwise, though we note that such values are not guaranteed.

%Taking a value $k_{sb}= 1.6 \times 10^{-12}$ m$^2$, as is consistent with the findings of \cite{cairns2025groundwater}, gives $K=1$ (see Appendix). 
%Values of $k_{sb}$ could range as low as $10^{-17}$ m$^2$ (giving $K\approx6 \times 10^{-5}$) \citep[\textit{e.g.}][]{gooch2016potential}, but are unlikely to exceed $10^{-12}$ m$^2$ (giving $K\approx6$). 
Finally, the parameter 
\begin{equation}
%\Sigma = \frac{\rho_i S_s (1-\xi) [H]^2}{\rho_w [m][t]}.
\Sigma = \frac{S_s [H]^2}{[m][t]}.
\end{equation} 
determines the relative size of exfiltration driven by compaction. 
Using a range from $S_s=1 \times 10^{-7}$ m$^{-1}$ to $1 \times 10^{-4}$ m$^{-1}$, as discussed above, leads to $\Sigma$ between 0.07 and 70.
%\cite{robel2023contemporary} use a range of $S_s$ from $1 \times 10^{-7}$ m$^{-1}$ to $1 \times 10^{-4}$ m$^{-1}$, which give $\Sigma$ between $5 \times 10^{-2}$ and $50$. 
%In this paper, we therefore consider values ranging from $S_s=0$ m$^{-1}$ ($\Sigma=0$) to $S_s=2.37 \times 10^{-5}$ m$^{-1}$ ($\Sigma = 12$). 

Equations \eqref{eq:massi*}, \eqref{eq:momih*} and \eqref{eq:hyd*2} are combined with non-dimensional analogues of the boundary conditions \eqref{eq:x0i}, \eqref{eq:Hixg}, \eqref{eq:stressxg}, \eqref{eq:hx0} and \eqref{eq:Nxg}, which may be found in Appendix A. 

\section{Steady ice sheet with weak groundwater flow}

\begin{figure}[t]
    \centering
    \includegraphics[width=0.9\linewidth]{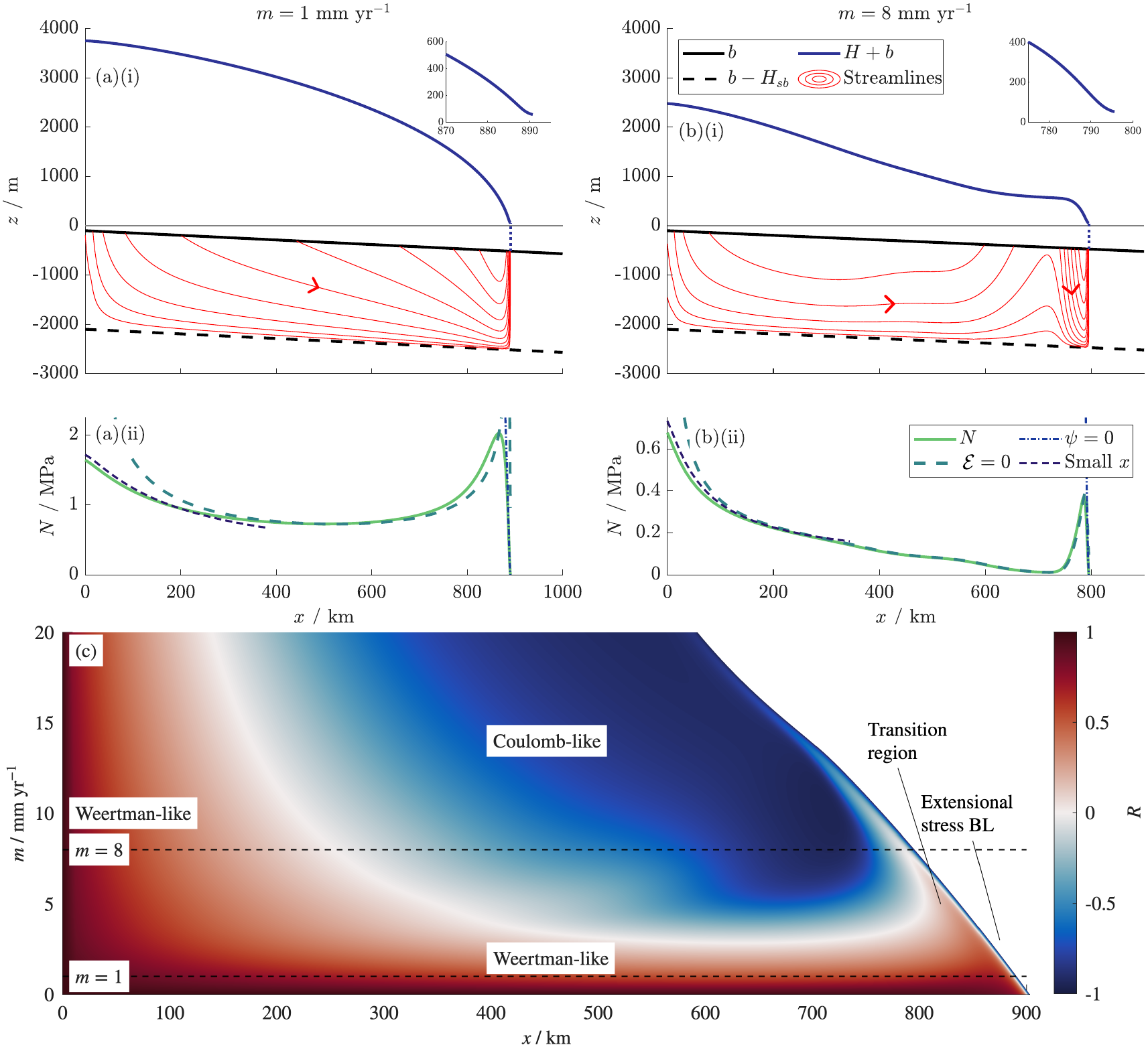}
    \caption{(i) Ice sheet profile and groundwater streamlines, and (ii) effective pressure $N$ for (a) $m=1$ mm yr$^{-1}$ (dimensionless $m=1$) and (b) $m=8$ mm yr$^{-1}$ (dimensionless $m=8$), and for $K \ll 1$, with uniform sedimentary basin thickness $H_{sb}=2000$ m. Inset panels show the ice profile near the grounding line. The effective pressure is compared to the $\mathcal{E}=0$ and $\psi=0$ approximations given by Equation \eqref{eq:NE0} and \eqref{eq:Npsi0} respectively, and the small-$x$ solution given by Equation \eqref{eq:Nnear0exact}. The bed is given by $b(x)  = (-100  - 230\times(x/500 \,\mathrm{km}))$m. 
    (c) Variation of the sliding law \eqref{eq:rc}  with $x$ as $m$ increases for $K \ll 1$. The colour corresponds to the value of  $R=(|u| - (\mathcal{C} N )^{1/m_W} ) /(|u| + (\mathcal{C} N )^{1/m_W} )$, so that red-purple ($0<R<1$) indicates the Weertman limit and blue ($-1<R<0$) indicates the Coulomb limit.}
    \label{fig:SS_K0}
\end{figure}

Firstly, to gain an intuition of the model, we consider the steady states in the limit $K\ll 1$, in which the sedimentary basin does not exert any effect on the subglacial effective pressure or ice thickness, although it is still possible to plot the limiting streamlines of groundwater flow. 
In the steady state, the compaction terms do not play a role, since these require time dependence by definition. 
(We obtain figures for the `$K\ll 1$' case by using the value $K=0$, although we refrain from referring to them as `$K=0$' as this would imply no groundwater flow occurs at all.)
%We note that although the feedback of the sedimentary basin is vanishingly small, groundwater flow can still occur in the sedimentary basin, and it is possible to determine the corresponding streamlines. 
% We consider solutions for the case of a linearly sloping bed $b(x) = (-100  - 230\times(x/500 \,\mathrm{km}))$m and, initially, for a uniform sedimentary basin thickness $H_{sb}$.

%This is equivalent to taking $K=0$, since the compaction terms do not appear in the steady state. 
\subsection{Ice sheet and effective pressure for weak groundwater flow}

% * Introductory sentence
Figure \ref{fig:SS_K0} shows examples of the steady-state ice sheet profile when $K \ll 1$, and how this profile varies with the melt rate $m$ via the dependence of the sliding law on the effective pressure $N$. The numerical method used to obtain these and subsequent solutions is described in Appendix C. 
Panels (a) and (b) show the ice sheet profile and effective pressure for the specific cases $m=1$ mm yr$^{-1}$ and $m=8$ mm yr$^{-1}$, whereas panel (c) shows how the behaviour of the sliding law varies as $m$ changes continuously. 
The effective pressure $N$ in panels (a)(ii) and (b)(ii) is compared with various approximate solutions, which are detailed in the next subsection.

In Figure \ref{fig:SS_K0}(a), the melt rate $m=1$ mm yr$^{-1}$ (dimensionless $m=1$) is relatively low. 
As a result, the effective pressure is sufficiently high that the sliding law \eqref{eq:rc} closely approximates the limiting case $\tau_b \approx {|u|}^{m_W}$ away from the grounding line, resulting in a `convex' ice sheet profile, which becomes steeper in the direction of flow. 

% * Near the grounding line, it's such and such
Near the grounding line, the sliding stress $\tau_b \to 0$, due to the condition \eqref{eq:Nxg} that $N\to 0$ as $x\to x_g$. 
This results in a boundary layer near $x_g$, where $\tau_b \approx \mathcal{C}N$ and the extensional stress terms in \eqref{eq:momih*} become significant. 
This boundary layer, which determines $x_g$, has been described by \citet{tsai2015marine}. 
In this boundary layer, seen in the inset panel in Figure \ref{fig:SS_K0}(a)(i), the ice profile ceases to be `convex' and flattens out following a point of inflection. 

Figure \ref{fig:SS_K0}(b) shows the steady state of the ice and hydrology at a higher melt rate $m=8$ mm yr$^{-1}$ (dimensionless $m=8$). 
Although the ice sheet in Figure \ref{fig:SS_K0}(b)(i) follows a similar `convex' shape to the $m=1$ mm yr$^{-1}$ case near the ice divide $x=0$, it assumes a much shallower shape nearer the grounding line before steepening again. 
As a result of this shallow portion, the ice sheet is thinner overall, reaching a maximum thickness of around 2500 m, rather than 4000 m as for $m=1$ mm yr$^{-1}$. 
Since the steady state ice flux from \eqref{eq:massi*} is simply the integral of the accumulation $a$, thinner ice must also be faster-flowing. 

The reason for this changed shape lies in the effective pressure $N$, seen in Figure \ref{fig:SS_K0}(b)(ii). 
When $m$ is increased, the overall size of $N$ away from the grounding line is decreased. %, following the $\mathcal{E}=0$ approximate solution \eqref{eq:NE0}. 
%However, the $\psi=0$ approximate solution \eqref{eq:Npsi0} still holds near the grounding line, meaning that the effective pressure continues to exhibit a maximum near $x_g$. 
In this case, $N$ is sufficiently low that the basal stress given by \eqref{eq:rc} behaves as $\tau_b \sim \mathcal{C}N$, rather than $\tau_b\sim{|u|}^{m_W}$, over much of the interior of the ice sheet. 
Since the basal friction balances the driving stress in \eqref{eq:momih*}, the reduced basal friction results in a lower $|\partial H / \partial x |$ and hence a shallower ice profile. 
However, near the ice divide $x=0$, the velocity becomes sufficiently low that the behaviour $\tau_b \approx {|u|}^{m_W}$ takes over once again, resulting in a reversion to the `convex' ice sheet profile. 

% * We had some more rigorous theory to expand this out but it's not especially insightful
There is also a retreat of the grounding line in Figure \ref{fig:SS_K0}(b)(i) compared to Figure \ref{fig:SS_K0}(a)(i), from around 920 km to 830 km. 
This is because increasing $m$  lowers the effective pressure close to $x_g$, hence reducing the basal friction in the extensional stress boundary layer of the ice (again visible in the inset panel). 
%This retreat is due to the sub-leading order effect of $m$ in the behaviour of $N$ near $x_g$, which reduces the effective pressure and hence the basal friction in the extensional stress boundary layer of the ice (again visible in the inset panel). 
As shown by \citet{tsai2015marine}, the reduced basal friction results in a higher ice flux across the grounding line for a given ice thickness, meaning the ice must reduce its thickness at the grounding line in response. 
This is achieved by grounding in shallower water (due to the flotation condition), and hence at a more retreated position. 

Figure \ref{fig:SS_K0}(c) shows how the behaviour of the sliding law \eqref{eq:rc} changes as $m$ is increased. For $m<3$ mm yr$^{-1}$, the sliding law is Weertman-like throughout the entire ice sheet, apart from in the extensional stress boundary layer, which is visible as a thin band of Coulomb-like behaviour near the grounding line. For $m$ between around 3 and 8 mm yr$^{-1}$, a Coulomb-like region emerges in the interior of the ice sheet, but is separated from the extensional stress boundary layer by a Weertman-like transition region caused by the local maximum in the effective pressure. For higher values of $m$, this maximum is lowered and the Coulomb-like behaviour dominates apart from near the origin, where $u \to 0$. The grounding line can also be seen to retreat as $m$ increases.

\subsection{Asymptotic structure of the effective pressure}

The behaviour of the effective pressure $N$ as $m$ changes can be explained by deriving various approximate solutions for $N$, which apply in different parts of the domain. 
These solutions are plotted alongside the effective pressure in Figures \ref{fig:SS_K0}(a)(ii) and Figures \ref{fig:SS_K0}(b)(ii). 

% Overview of the equation
Integrating the steady state of Equation \eqref{eq:hyd*2} subject to the no-flux condition \eqref{eq:hx0} gives for a general $K$
\begin{equation}\label{eq:NSS}
\left(h(N) +KH_{sb}\right)  \frac{\partial}{\partial x}{} \left( (1-\delta)H - \mathcal{E} N +  b \right)  + mx =0.
\end{equation}
The approximate solutions of this equation emerge from the fact that $\mathcal{E}$ is relatively small. 
%Since $\mathcal{E}$ is relatively small, the solution for $N(x)$ follows two approximate solutions in different regions. 
Away from the grounding line $x=x_g$ and from $x=0$, the solution is roughly given by the case when $\mathcal{E}=0$, which for $K \ll1$ reads
\begin{equation}\label{eq:NE0}
% h(N) \sim mx \left[ -   \frac{\partial}{\partial x}{} \left( (1-\delta)H + b \right)\right]^{-1} - KH_{sb}, \quad \text{for} \quad x\gg \sqrt{\frac{\mathcal{E}}{m}}, \quad x_g-x\gg \frac{\mathcal{E}}{mx_g}.
h(N) \sim mx \left[ -   \frac{\partial}{\partial x}{} \left( (1-\delta)H + b \right)\right]^{-1}, \quad \text{for} \quad x\gg \sqrt{\frac{\mathcal{E}}{m}}, \quad x_g-x\gg \frac{\mathcal{E}}{mx_g}.
\end{equation}
This equation results from the fact that the effective pressure contribution to the hydraulic potential is much smaller than that due to the ice thickness $H$ and bed elevation $b$. 
Since $h(N)$ is a decreasing function (we take $h(N) =1/N$ in particular), Equation \eqref{eq:NE0} implies that increasing $m$ decreases $N$ away from the grounding line. 
% In particular, for $h(N) =1/N$ we have 
% \begin{equation}\label{eq:NE0K2}
% N\sim \frac{-\frac{\partial}{\partial x}{} \left( (1-\delta)H + b \right)}{mx + KH_{sb}  \frac{\partial}{\partial x}{} \left( (1-\delta)H + b \right)}, \quad \text{for} \quad x\gg \sqrt{\frac{\mathcal{E}}{m}}, \quad x_g-x\gg \frac{\mathcal{E}}{mx_g}.
% \end{equation}
%Equation \eqref{eq:NE0} breaks down for small $m$ and/or large $K$, as when $h$ becomes small the large effective pressure $N$ can no longer be neglected in the hydraulic potential. We explore the limit of large $K$ in Section 4.1. 
Equation \eqref{eq:NE0} breaks down for small $m$, as when $h$ becomes small the large effective pressure $N$ can no longer be neglected in the hydraulic potential. %We explore the limit of large $K$ in Section 4.1. 

Near the grounding line, the approximate solution \eqref{eq:NE0} fails to satisfy the $N=0$ boundary condition. 
Provided that $h(N) \to \infty$ as $N \to 0$, as is the case for our choice of $h(N)$, the solution of Equation \eqref{eq:NSS} is approximately given by equating the hydraulic potential $\psi$ to zero, so that
\begin{equation}\label{eq:Npsi0}
 N \sim \frac1{\mathcal{E}} \left((1-\delta)H +b\right) \quad \text{for} \quad x_g-x\ll \frac{\mathcal{E}}{mx_g}.
\end{equation}
(If $h$ is finite but large, Equation \eqref{eq:Npsi0} overestimates $N$ by an amount scaling roughly with $1/h(x_g,t)$.) 
The effective pressure in Figure \ref{fig:SS_K0}(a)(ii) exhibits a local maximum near the grounding line as the solution transitions from Equation \eqref{eq:NE0} to Equation \eqref{eq:Npsi0}. 
The local behaviour \eqref{eq:Npsi0} of $N$ near the grounding line can be thought of mathematically as a boundary layer associated with the small parameter $\mathcal{E}$, but should not be confused with the extensional stress boundary layer in the ice sheet associated with the small parameter $\varepsilon$. 

The extensional stress boundary layer has thickness $O({\varepsilon}^{n/(n+2)})$, and therefore the effective pressure within this layer is described by the $\psi=0$ approximate solution above provided ${\varepsilon}^{n/(n+2)} \ll \mathcal{E}/ mx_g$. 
For the values in Table 2 (see Appendix), ${\varepsilon}^{n/(n+2)} \approx 0.014$. 
In this case, the sliding stress is effectively proportional to the hydraulic potential,
\begin{equation}
\tau_b \sim \mathcal{C}N \sim \frac{\mathcal{C}}{\mathcal{E}} ((1-\delta)H +b),
\end{equation}
which is exactly the case considered by \citet{tsai2015marine}. However, if the boundary layer thicknesses are comparable, the outer behaviour of $N$ and in particular melt rate $m$ can influence the extensional stress boundary layer, and hence the grounding line position. 

%Because the grounding line position of the ice sheet is determined by the extensional stress boundary layer, in which the sliding stress $\tau_b\sim \mathcal{C}N$, Equation \eqref{eq:Npsi0} suggests that the melt rate $m$ and sedimentary basin do not affect the steady grounding line position to leading order, although they may have a small effect via sub-leading order terms. 

% * Transition region
\subsubsection{Transition region}
The $\mathcal{E}=0$ and $\mathcal{\psi}=0$ approximate solutions are connected by a transition region, which we may characterise, focusing on the case where the ice obeys the Weertman sliding law throughout this region. 
This enables us to explain the existence of the local maximum in $N$ near the grounding line seen in Figure \ref{fig:SS_K0}, and determine how this maximum scales with $m$. 
% *why not coulomb
Assuming we are outside of the extensional stress boundary layer, the steady state ice momentum equation \eqref{eq:momih*} is therefore approximately
\begin{equation}\label{eq:mom_reduced}
\left(\frac{a x}{H}\right)^{m_W} = - H \left(\frac{\partial H}{\partial x} + \frac{\partial b}{\partial x} \right).
\end{equation}
We rescale variables as
\begin{multline}\label{eq:trans_scales}
x-x_g = \frac{\mathcal{E}}{mx_g}\xi, \quad N = (1-\delta)\left(\frac{a^{m_W} {x_g}^{m_W-1}}{m \mathcal{E}^{m_W+1}} \right)^{1/(m_W+2)} \hat{N}, \\ H = \left(\frac{a^{m_W} \mathcal{E} {x_g}^{m_W-1}}{m}  \right)^{1/(m_W+2)} \hat{H},
\end{multline}
After this scaling, Equations \eqref{eq:mom_reduced} and \eqref{eq:NSS} become respectively the coupled equations for $\hat{H}$ and $\hat{N}$ 
% \begin{equation}
% \frac{\partial \hat{H}}{\partial \xi} =  \frac1{\hat{H}^{m_W+1}} -\hat{\beta} ,
% \end{equation}
% and the equation for $N$ is
% \begin{equation}
%   \frac{\partial \hat{N}}{\partial \xi}  =\frac{\partial \hat{H}}{\partial \xi} + \frac{\hat{\beta}}{1-\delta}  -  \hat{N} ,
% \end{equation}
% where
% \begin{equation}
% \hat{\beta} = \left(\frac{\mathcal{E}^{m_W+1}}{a^{m_W} m^{m_W+3} {x_g}^{2m_W+3} } \right)^{1/(m_W+2)} \frac{\partial b}{\partial x} \ll 1,
% \end{equation}
% is negligibly small.
% After ignoring these terms, we have  
\begin{equation}\label{eq:transH}
\frac{\partial \hat{H}}{\partial \xi} =  \frac1{\hat{H}^{m_W+1}} ,
\end{equation}
\begin{equation}\label{eq:transN}
 % \frac{\partial \hat{N}}{\partial \xi}  =\frac1{\hat{H}^{m_W+1}}    -  \frac{\hat{N}}{1+\hat{\kappa} \hat{N}},
 \frac{\partial \hat{N}}{\partial \xi}  =\frac1{\hat{H}^{m_W+1}}    - \hat{N},
\end{equation}
% where 
% \begin{equation}
% \hat{\kappa} =(1-\delta)\left(\frac{a^{m_W} m {x_g}^{m_W-1}}{\mathcal{E}^{m_W+1}} \right)^{1/(m_W+2)} KH_{sb}(x_g),
% \end{equation}
assuming that $\left(\mathcal{E}^{m_W+1}/a^{m_W} m^{m_W+1} {x_g}^{2m_W+1}  \right)^{1/(m_W+2)} \partial b /\partial x \ll 1 $ so that the bedslope terms are negligible, which is true for our chosen parameter values. 

%As described in the Appendix, the boundary conditions \eqref{eq:Hixg*} and \eqref{eq:Nxg*} give
The boundary conditions \eqref{eq:Hixg} and \eqref{eq:Nxg} give (see Appendix equations \eqref{eq:Hixg*} and \eqref{eq:Nxg*})
\begin{equation}
\hat{H}=\hat{H}_g, \quad \hat{N}=0 \quad \text{at} \quad \xi=0,
\end{equation}
where 
\begin{equation}\label{eq:Hg}
\hat{H}_g = \big(m/a^{m_W} \mathcal{E} {x_g}^{m_W-1}  \big)^{1/(m_W+2)} b(x_g)/(1-\delta)
\end{equation} is the scaled ice thickness required for flotation.
Matching to the $\mathcal{E}=0$ approximate solution \eqref{eq:NE0} requires that 
\begin{equation}\label{eq:match_trans}
% \hat{N} \sim \frac{1}{\hat{H}^{m_W+1} - \hat{\kappa} } \quad \text{as} \quad \xi\to\infty.
\hat{N} \sim \frac{1}{\hat{H}^{m_W+1}} \quad \text{as} \quad \xi\to\infty.
\end{equation}
The $\mathcal{\psi}=0$ approximate solution \eqref{eq:Npsi0}, which after scaling gives $\hat{N} \sim \hat{H}-\hat{H}_g$ as $\xi\to0$, follows automatically from the equations and boundary conditions by Taylor expanding $\hat{N}-\hat{H}$ about $\xi =0$.  

\begin{figure}
    \centering
    \includegraphics[width=1\linewidth]{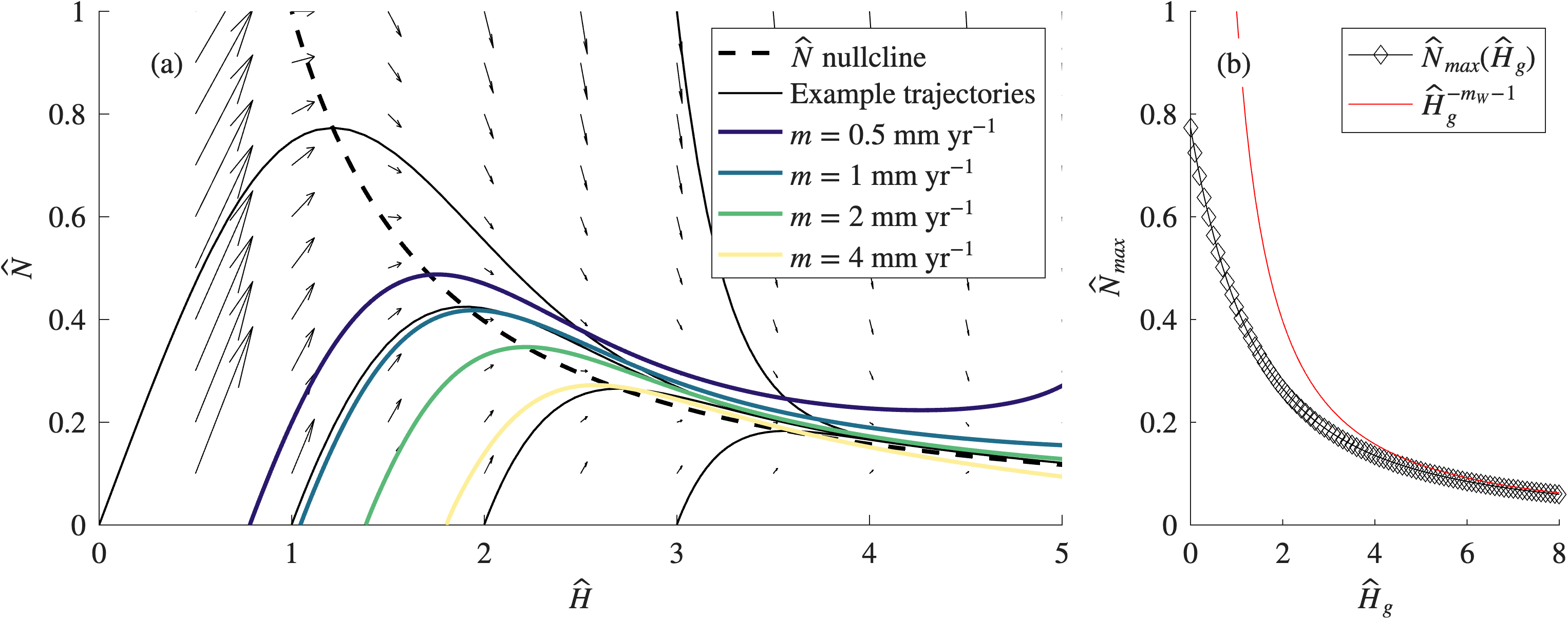}
    \caption{(a) Phase portrait of Equations \eqref{eq:transH} and \eqref{eq:transN}, including scaled plots of the numerical  solution of the full problem for selected $m$ with $\hat{\kappa}=0$. (b) Maximum value of $\hat{N}$ attained for varying initial conditions $\hat{H}_g$, compared with the asymptotic behaviour $\hat{H}_g^{-m_W-1}$.}
    \label{fig:trans_K0}
\end{figure}

% * Describe the trajectories
Figure \ref{fig:trans_K0}(a) shows the phase portrait of Equations \eqref{eq:transH} and \eqref{eq:transN}. Trajectories initially behave as $\hat{N} \sim \hat{H}-\hat{H}_g$ (the $\psi=0$ approximate solution near the grounding line), until they cross the $N$-nullcline and reach a maximum $\hat{N}_{max}$, whereafter the trajectories decay towards the nullcline. 
This nullcline coincides with the $\mathcal{E}=0$ approximate solution, ensuring the matching condition \eqref{eq:match_trans} is satisfied. 
Since trajectories must begin on the $\hat{N}=0$ axis, the trajectory from $\hat{H}=0$, $\hat{N}=0$ acts as an upper bound for physically relevant solutions.  
In Figure \ref{fig:trans_K0}(b),  $\hat{N}_{max}$ is plotted as a function of the initial condition ${\hat H}_g$, which behaves asymptotically as $\hat{N}_{max} \sim \hat{H}_g^{-m_W-1}$ for large $\hat{H}_g$.

Scaled solutions of the full steady-state problem (Equation \eqref{eq:NSS}) are plotted for selected $m$, and generally agree well with the phase trajectories. 
The theory begins to break down for small $m$ as the transition layer thickness $\mathcal{E}/m$ becomes large, and for large $m$ as the assumption of a Weertman sliding law breaks down. 
Overall, a higher $m$ corresponds to a higher ${\hat H}_g$ and hence a lower $\hat{N}_{max}$, leading to an lower maximum $N$ overall. 
More specifically, the scalings \eqref{eq:trans_scales} give the (local) maximum  of $N$ as
\begin{equation}
N_{max} =(1-\delta)\left(\frac{a^{m_W} {x_g}^{m_W-1}}{m \mathcal{E}^{m_W+1}} \right)^{1/(m_W+2)} \hat{N}_{max} \left(  \left(\frac{m} {a^{m_W} \mathcal{E} {x_g}^{m_W-1}} \right)^{1/(m_W+2)} \frac{b(x_g)}{1-\delta}\right)
\end{equation}
Since $\hat{N}_{max} \sim \hat{H}_g^{-m_W-1}$ for large $\hat{H}_g$, we have that
\begin{multline}
N_{max} \sim(1-\delta)\left(\frac{a^{m_W} {x_g}^{m_W-1}}{m \mathcal{E}^{m_W+1}} \right)^{1/(m_W+2)}  \left(\frac{a^{m_W} \mathcal{E} {x_g}^{m_W-1}}{m} \right)^{(m_W+1)/(m_W+2)} \left( \frac{b(x_g)}{1-\delta}\right)^{-m_W-1} \\
\sim\frac{a^{m_W} {x_g}^{m_W-1} (1-\delta)^{m_W+2} }{m b(x_g)^{m_W+1}}  ,
\end{multline}
so the local maximum of $N$ decays as $m^{-1}$, similarly to the $\mathcal{E}=0$ approximate solution.  
%which decreases as $m$ increases, assuming that the $m^{-1}$ factor dominates compared to the effects of grounding line movement. 

\subsubsection{Approximate solution for small $x$}
The $\mathcal{E}=0$ solution also breaks down near $x=0$, because both the ice thickness gradient and overall water flux $mx$ in Equation \eqref{eq:NSS} become small. 
In this region, the ice sliding law again approximates the Weertman sliding law with negligible extensional stress, so that we can approximate (from Equation \eqref{eq:mom_reduced})
\begin{equation}
(1-\delta)\frac{\partial H}{\partial x} + \frac{\mathrm{d} b}{\mathrm{d} x} \approx\frac{\partial H}{\partial x} + \frac{\mathrm{d} b}{\mathrm{d} x}
\approx - \frac{(ax)^{m_W}}{H^{m_W+1}}.
\end{equation}
Equation \eqref{eq:NSS} then becomes approximately
\begin{equation}\label{eq:NSS_near0}
% - \frac{(ax)^{m_W}}{H^{m_W+1}} - \mathcal{E} \frac{\partial N}{\partial x}   + \frac{mx N}{1 + K H_{sb} N} =0.
- \frac{(ax)^{m_W}}{H^{m_W+1}} - \mathcal{E} \frac{\partial N}{\partial x}   + mx N =0.
\end{equation}
Equation \eqref{eq:NSS_near0} can be solved explicitly to obtain (after re-scaling)
\begin{align}\label{eq:Nnear0exact}
N &= \frac{a^{m_W}}{\mathcal{E}H(0)^{m_W+1}} \int_x^\infty {x'}^{m_W} \exp\left( -\frac{m({x'}^2-x^2)}{2\mathcal{E}} \right) \, \mathrm{d}x'\\
&= \frac{(2\mathcal{E})^{(m_W -1)/2} a^{m_W}}{H(0)^{m_W+1} m^{(m_W +1)/2}} \exp\left( \frac{mx^2}{2\mathcal{E}} \right) \Gamma\left(\frac{m_W+1}2,\frac{mx^2}{2\mathcal{E}}\right),
\end{align}
where $\Gamma$ is the upper incomplete gamma function. 
Equation \eqref{eq:Nnear0exact} approximates $N$ for $x = O(\sqrt{\mathcal{E}/m})$. 
As $m$ increases, $N$ scales with $m^{-(m_W+1)/2}$, decreasing more slowly than the $\mathcal{E}=0$ approximate solution \eqref{eq:NE0}. This is why the effective pressure in Figure \ref{fig:SS_K0}(b)(ii) is relatively high near the origin compared to the rest of the domain. 
% For K>0, we can analyse these equations

% If we rescale the variables as 
% \begin{equation}\label{eq:scalings_smallx}
% x = \sqrt{\frac{\mathcal{E}}{m}} \tilde{x}, \quad N = \frac{a^{m_W} \mathcal{E}^{(m_W-1)/2}  }{H(0)^{m_W+1} m^{(m_W+1)/2} } \tilde{N},
% \end{equation}
% we obtain at leading order a single equation for $\tilde{N}$, decoupled from $\tilde{H}$,
% \begin{equation}\label{eq:near0_scaled}
% \frac{\partial \tilde N}{\partial \tilde x} = - \tilde{x}^{m_W}     + \tilde{x} \tilde{N} ,
% \end{equation}
% since $H=H(0)$ and $H_{sb}=H_{sb}(0)$ are constant to leading order. 
% Matching with the $\mathcal{E}=0$ approximate solution \eqref{eq:NE0} requires that
% \begin{equation}\label{eq:asymp_near0}
% \tilde N  \sim \tilde{x}^{m_W-1} \quad \text{as} \quad \tilde{x} \to \infty.
% \end{equation}
% The solution of Equation \eqref{eq:near0_scaled} subject to \eqref{eq:asymp_near0} gives an approximate solution for $x = O(\sqrt{\mathcal{E}/m})$. 
% As $m$ increases, $N$ scales with $m^{-(m_W+1)/2}$, decreasing more slowly than the $\mathcal{E}=0$ approximate solution, which scales with $m^{-1}$ for $K\ll 1$. This is why the effective pressure in Figure \ref{fig:SS_K0}(b)(ii) is relatively high near the origin compared to the rest of the domain. 

% * Exact solution
%\subsubsection{Exact solution}
%For the particular case where $K=0$, 

In the case where $K\ll1$, Equation \eqref{eq:NSS} in fact has an exact solution in terms of $H$ and $b$, given by
\begin{equation}\label{eq:Nexact}
N(x) = - \frac1{\mathcal{E}}\int_x^{x_g} \left( (1-\delta)\frac{\partial H}{\partial x}(x') + \frac{\partial b}{\partial x}(x') \right) \exp \left( -\frac{m ({x'}^2-x^2)}{2 \mathcal{E}} \right) \,\mathrm{d}x'.
\end{equation}
The $\mathcal{E}=0$ approximate solution \eqref{eq:NE0} may be recovered from Equation \eqref{eq:Nexact} using Laplace's method, assuming the appropriate constraints on $x$. 
Similarly, the small-$x$ solution \eqref{eq:asymp_near0} and the transition region can also be recovered at leading order by applying the required approximations.

\subsection{Streamlines of groundwater flow}

We now consider the groundwater flow, exfiltration and infiltration for $K \ll 1$. 
Equation \eqref{eq:qE}, after non-dimensionalising, gives the exfiltration in the steady state as
\begin{equation}\label{eq:qE2}
%q_E =  K   \frac{\partial}{\partial x}{} \left[ H_{sb}   \frac{\partial}{\partial x}{} \left( H - \mathcal{E} N +  \frac{b}{1-\delta} \right) \right].
q_E =  K   \frac{\partial}{\partial x}{} \left[ H_{sb}   \frac{\partial}{\partial x}{} \left( (1-\delta) H - \mathcal{E} N +  b \right) \right].
\end{equation}
Upon substituting in Equation \eqref{eq:NSS} for the hydraulic potential gradient, we have 
\begin{equation}\label{eq:qEalt}
\begin{split}
q_E &=  - K   \frac{\partial}{\partial x}{} \left[ H_{sb}\frac{  mx }{h+K H_{sb} }\right]  \\
&= -\frac{KmH_{sb}}{h+K H_{sb}}-\frac{Kmx}{h^2(h+K H_{sb})^2}  \frac{\partial}{\partial x}{}\left(\frac{H_{sb}}{h} \right) .
\end{split}
\end{equation}
%Since $h$ and $H_{sb}$ are positive, 
Equation \eqref{eq:qEalt} implies that a necessary (but insufficient) condition for $q_E \geq 0$ (\textit{i.e.} exfiltration) is that 
\begin{equation}\label{eq:qEonlyif}
  \frac{\partial}{\partial x}{}\left(\frac{H_{sb}}{h(N)}\right) <0.
\end{equation}
That is, the ratio of the groundwater flux through the sedimentary basin to that through the cavity sheet system is decreasing. 
%Since $KH_{sb} N = KH_{sb}/h$ is the ratio of the groundwater flux through the sedimentary basin to that through the cavity sheet system, Equation \eqref{eq:qEonlyif} implies that exfiltration occurs only if this ratio is decreasing. 
This condition could be satisfied through the sedimentary basin thickness $H_{sb}$ decreasing in the direction of flow, as considered previously by \citet{cairns2025groundwater}. 
The other possibility is that $h(N)$ is increasing, \textit{i.e.} $N$ is decreasing. 
Although this is generally expected over most of the domain (as the water flux through the domain increases due to melting), this constraint can provide insight into where exfiltration occurs, as seen below. 
%, although this constraint is less powerful in some cases than others. 

The streamlines of groundwater flow in the limit $K \ll 1$ with constant $H_{sb}=2000$ m are plotted in Figure \ref{fig:SS_K0}(a)(i) and (b)(i). 
%We repeat the two cases from Figure \ref{fig:SS_K0} of $m=10$ mm yr$^{-1}$ and $m=80$ mm yr$^{-1}$ in Figure \ref{fig:SS_K0}(a) and (b) respectively. 
In the $m=1$ mm yr$^{-1}$ case (Figure \ref{fig:SS_K0}(a)(i)), groundwater infiltrates into the sedimentary basin almost everywhere, apart from a small region of exfiltration near the grounding line. 
This is because the condition \eqref{eq:qEonlyif} implies for constant $H_{sb}$ that exfiltration can only occur where $\partial N / \partial x<0$.  
Since for $x<500$ km the steepening gradient of the ice overburden drives infiltration, the region of exfiltration is limited to where $N$ decreases in the transition region near the grounding line.

In the $m=8$ mm yr$^{-1}$ case (Figure \ref{fig:SS_K0}(b)(i)), however, this pattern is modified by an intermediate region of exfiltration, occurring where the flattening of the ice surface leads to decreasing $N$ following the $\mathcal{E}=0$ approximate solution \eqref{eq:NE0}. 
If we increased $K$, we would expect this exfiltration to further reinforce the lowering of $N$ and consequent flattening of the ice. 

% * I don't think this explanation is correct! the change in sign is actaully due to a turning point in the outer ice solution, which is barely visible but still governed by the E=0 approx solution.
Another way to think of $q_E$, alternatively to Equation \eqref{eq:qEalt}, is in terms of the hydraulic potential $\psi$. 
For constant $H_{sb}$, $q_E$ is proportional to $\partial^2 \psi /\partial x^2$ by Equation \eqref{eq:qE2}. 
Away from the grounding line, $\psi$ is dominated by the ice overburden $H$, so that the `convex' ice profile in Figure \ref{fig:SS_K0}(a)(i) leads to infiltration. 
Exfiltration only occurs near the grounding line, where the $\psi\approx0$ approximate solution \eqref{eq:Npsi0} and the extensional stress boundary layer lead to a change in sign of $\partial^2 \psi /\partial x^2$. 
In Figure \ref{fig:SS_K0}(b)(i), however, the more complex shape of the ice means $\partial^2 \psi /\partial x^2$ temporarily changes sign away from the grounding line, leading to an additional region of exfiltration.

\section{Steady ice sheet with strong groundwater flow}

We now consider the steady states of the model when exfiltration and infiltration are strong enough to impact the hydrology and ice sheet, \textit{i.e.} $K=O(1)$. 
As we have mentioned above, realistic values of $K$ include both small values and $O(1)$ values up to around $K\approx6$. 
We begin with the case of of a uniform sedimentary basin thickness $H_{sb}=2000$ m for simplicity, before considering the additional effect of variation in $H_{sb}$. 
% Throughout this section, we refer to the cases $K \ll 1$ and $K=0$ interchangeably, although they are subtly different in that the former allows streamlines to be plotted, whereas the latter implies no groundwater flow at all. 

\subsection{Low melt rate}

% Fewer streamlines. add edge. use dimensional values for consistency
\begin{figure}[t]
    \centering
    \includegraphics[width=.75\linewidth]{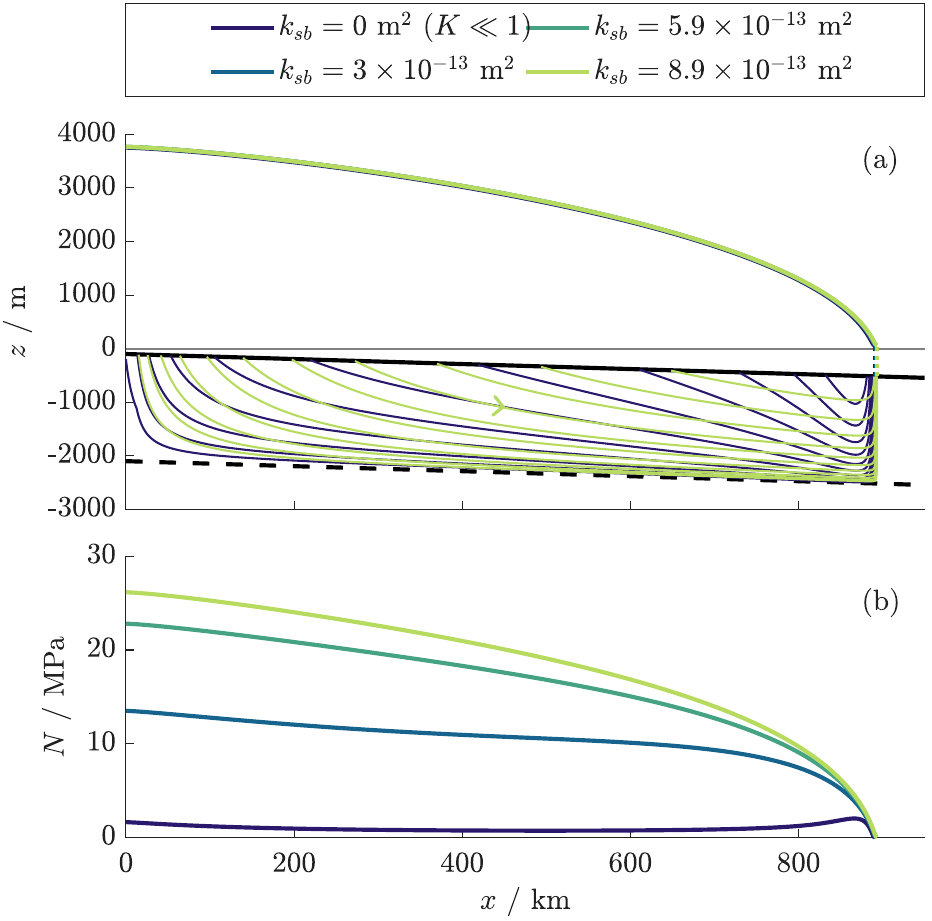}
    \caption{(a) Ice sheet profile and sedimentary basins, and (b) effective pressure for a relatively low melt rate $m=1$ mm yr$^{-1}$ (dimensionless $m=1$), as in Figure \ref{fig:SS_K0}(a), but with various values of $k_{sb}$ corresponding to $K\ll1$ and $K=2,4,6$ respectively, and a uniform sedimentary basin depth $H_{sb} = 2000$ m. The streamlines are only plotted for $k_{sb} =0$ m$^2$ and the maximum value of $k_{sb}$.}
    \label{fig:SS_varyK_m1}
\end{figure}

We first consider the case where $m=1$ mm yr$^{-1}$ (dimensionless $m=1$) is sufficiently small that the ice sheet is largely unaffected by subglacial hydrology ($m=1$), as seen in Figure \ref{fig:SS_K0}(a). 
This initial case enables us to observe the effect of groundwater on the effective pressure $N$ with the ice sheet held roughly constant. 
Figure \ref{fig:SS_varyK_m1}(a) shows the resulting ice sheet and streamlines, and Figure \ref{fig:SS_varyK_m1}(b) shows the effective pressure. 

In the case where the ice sheet is largely unaffected by subglacial hydrology, the main outcome of including a sedimentary basin  is to increase the effective pressure $N$, with a greater increase for higher $K$. 
This increase in $N$ can be seen from the $\mathcal{E}=0$ approximate solution \eqref{eq:NE0}, which for nonzero $K$ is given by
\begin{equation}\label{eq:NE0K}
h(N) \sim mx \left[ -   \frac{\partial}{\partial x}{} \left( (1-\delta)H + b \right)\right]^{-1} - KH_{sb}, \quad \text{for} \quad x\gg \sqrt{\frac{\mathcal{E}}{m}}, \quad x_g-x\gg \frac{\mathcal{E}}{mx_g}.
\end{equation}
In particular, for $h(N) =1/N$ we have 
\begin{equation}\label{eq:NE0K2}
N\sim \frac{-\frac{\partial}{\partial x}{} \left( (1-\delta)H + b \right)}{mx + KH_{sb}  \frac{\partial}{\partial x}{} \left( (1-\delta)H + b \right)}, \quad \text{for} \quad x\gg \sqrt{\frac{\mathcal{E}}{m}}, \quad x_g-x\gg \frac{\mathcal{E}}{mx_g}.
\end{equation}
The increase in $N$ is a consequence of the fact that the total meltwater flux $mx$ must be shared between the sedimentary basin and the cavity sheet, leading to a lower effective water thickness $h$. 

To better understand the results in Figure \ref{fig:SS_varyK_m1}, we can consider the limit $K \to \infty$, in which the sedimentary basin dominates subglacial water transport. 
In this case, Equation \eqref{eq:qEalt} gives
\begin{equation}
q_E \sim -m + \frac{1}{K}\frac{\partial}{\partial x} \left(\frac{mxh}{H_{sb}} \right) + O\left(\frac{1}{K^2}\right),
\end{equation}
so that the infiltration is simply given by the melt rate at leading order, except in a small region where $h=O(K)$ near the grounding line. 
In this case, the $\mathcal{E}=0$ approximate solution \eqref{eq:NE0K} away from the grounding line breaks down. 
Instead, the horizontal groundwater flux is determined by the melt rate, and the effective pressure is determined by the required hydraulic gradient, so Equation \eqref{eq:NSS} gives
\begin{equation}
KH_{sb}\frac{\partial}{\partial x}{} \left( (1-\delta)H - \mathcal{E} N +  b \right)  + mx \approx0,
\end{equation}
implying that the hydraulic gradient $\partial \psi /\partial x$ is small. 
Hence, for the higher values of $k_{sb}$ in Figure \ref{fig:SS_varyK_m1}, $N$ roughly follows the shape of the ice surface. 
In terms of the asymptotic regions considered in Section 3.2, we can observe that the transition region between the $\mathcal{E}=0$ and $\psi=0$ approximate solutions grows in size with increasing $K$, as does the size of $N$ within this region, until this region comes to dominate the whole domain so that the $\mathcal{E}=0$ approximate solution no longer applies. 
We consider the impact of the sedimentary basin on the transition region and small-$x$ approximate solution below.
%We therefore have 
% \begin{equation}
% N \sim \frac1{\mathcal{E}} \left((1-\delta)H +b - \int_x^{x_g} \frac{mx'}{K H_{sb}}\, \mathrm{d}x'\right),
% % * weird stuff at the boundary?
% \end{equation}
% which resembles the `$\psi=0$' approximate solution \eqref{eq:Npsi0}. 

For smaller $k_{sb}$, the $\mathcal{E}=0$ approximate solution \eqref{eq:NE0K2} suggests that the sedimentary basin has a larger effect when the distance from the ice divide $x$ is small, or the hydraulic potential gradient is steep. 
When $m=1$ mm yr$^{-1}$, the hydraulic potential gradient becomes steeper with increasing $x$, so that these competing effects even out to a relatively uniform effect of varying $K$. 
Near the grounding line, $N$ follows the $\psi=0$ approximate solution \eqref{eq:Npsi0} independent of $K$, so that the sedimentary basin has little effect on the extensional stress boundary layer of the ice and resulting grounding line position $x_g$. 

For the nonzero $k_{sb}$ considered in Figure \ref{fig:SS_varyK_m1}, $N$ attains very large values ($>10$ MPa). 
This is partly because we have used values of $k_{sb}$ at the higher end of what is likely to be physically realistic. 
However, it is worth noting that in reality, at high effective pressures the ice would likely infiltrate the upper layers of sediment to form a `frozen fringe', significantly reducing the permeability \citep{rempel2008theory}, which we discuss further in Section 6.
% For the nonzero $k_{sb}$ considered in this case, $N$ attains unphysically large values. 

\subsection{High melt rate}

\begin{figure}[p]
    \centering
    \includegraphics[width=0.9\linewidth]{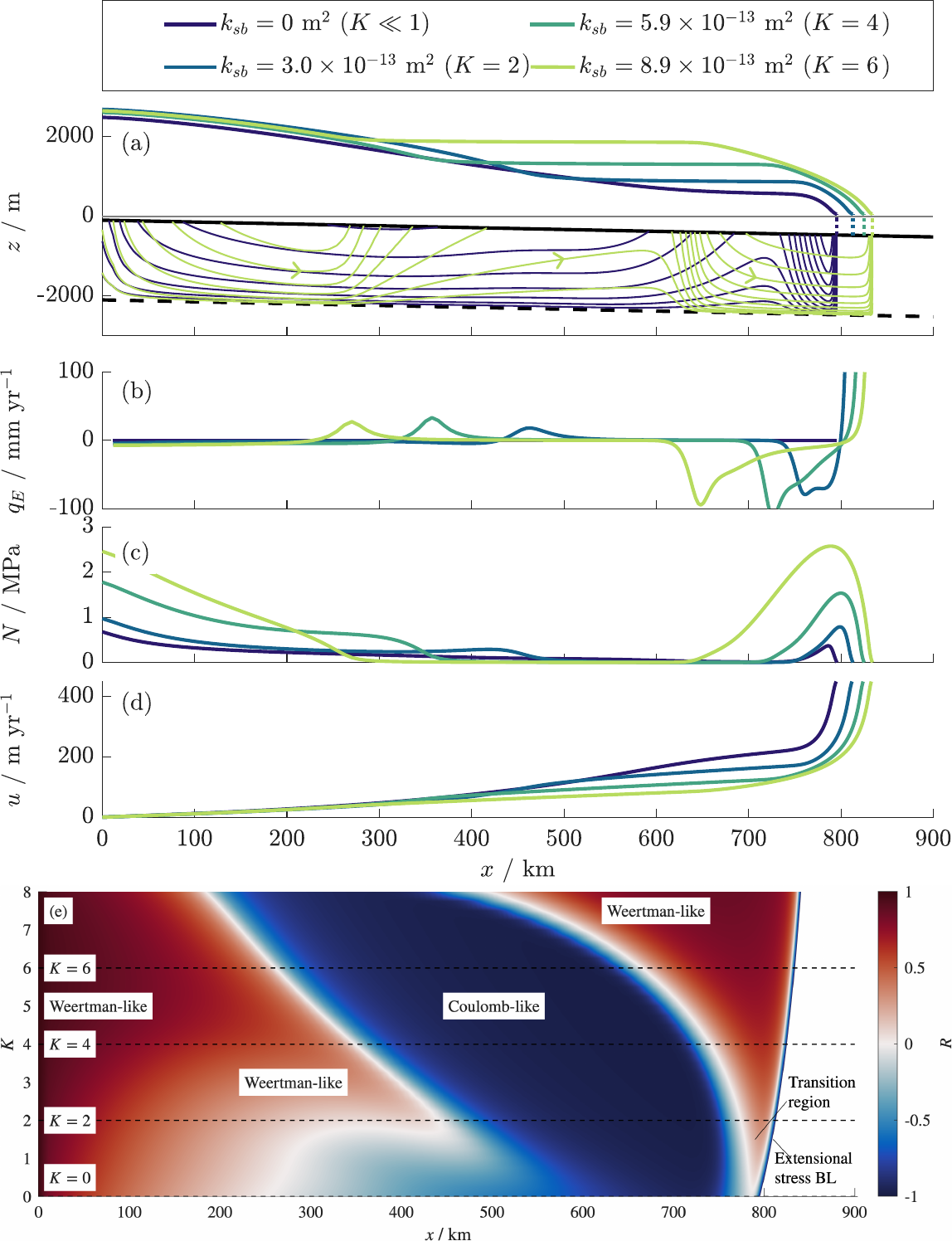}
    \caption{(a) Ice sheet profile and sedimentary basin streamlines, (b) exfiltration $q_E$, (c) effective pressure, and (d) ice velocity for a higher melt rate $m=8$ mm yr$^{-1}$ (dimensionless $m=8$), as in Figure \ref{fig:SS_K0}(b). The various values of $k_{sb}$, the sedimentary basin depth $H_{sb}$, and the plotted streamlines are as in Figure \ref{fig:SS_varyK_m1}. (e) Variation of the sliding law \eqref{eq:rc} with $x$ as $K$ increases for $m=8$ mm yr$^{-1}$, where red regions are Weertman-like and blue regions are Coulomb-like as in Figure \ref{fig:SS_K0}(c).}
    \label{fig:SS_varyK_m8}
\end{figure}

Figure \ref{fig:SS_varyK_m8} shows the effect of strong groundwater flow on the steady state ice sheet and hydrology for the higher melt rate $m=8$ mm yr$^{-1}$ (dimensionless $m=8$), where the hydrology affects the ice profile, as seen in Figure \ref{fig:SS_K0}(b). 

% * Can we exlpain the migration of this flat bit?
When the groundwater flow is weak, the low effective pressure results in a region of the ice with small surface slope, with a localised region of groundwater exfiltration slightly upstream. 
For strong groundwater flow, this exfiltration causes the expansion of the region of low effective pressure further upstream, in turn causing the region of exfiltration to move upstream, as seen in Figure \ref{fig:SS_varyK_m8}(b). 
The corresponding change in the effective pressure is seen in Figure \ref{fig:SS_varyK_m8}(c). 
Away from this region of low effective pressure, the presence of a more permeable sedimentary basin increases $N$, with the effect strongest where the hydraulic gradient (\textit{i.e.} the ice sheet profile) is steepest, near the ice divide and grounding line. 
The increase in $N$ in these regions can again be understood by considering the transition region and small-$x$ approximate solutions, which we address below. 

The effect on the ice sheet, along with the groundwater flow streamlines, is shown in Figure \ref{fig:SS_varyK_m8}(a), with the variation of the sliding law shown in Figure \ref{fig:SS_varyK_m8}(e). 
As the region of low effective pressure moves upstream with increasing $K$, the corresponding Coulomb-like region where the ice surface is relatively flat follows. 
The increase in effective pressure near the grounding line leads to a larger Weertman-like region where the basal friction is high, and the ice profile relatively steep as a result. 
The larger steep region leads to a larger ice thickness overall, as ice accumulates behind the region of increased friction. 
Since the ice flux is given by $Hu=ax$ in the steady state, the thickening of the ice sheet when $K$ is increased results in a reduced ice velocity, as shown in \ref{fig:SS_varyK_m8}(d). 
%In Figure(d), the ice velocity $u$ is plotted for various $K$, showing a substantial reduction as $K$ is increased. 

Finally, we can observe an advance of the grounding line as $K$ is increased. 
This effect emerges because increasing $K$ raises the effective pressure close to the grounding line, similarly to if $m$ were decreased. 
The increased friction results in a lower ice flux for the same ice thickness, so that the ice is forced to ground in deeper water and hence at a higher $x$.

% The implications of the E=0 approximate solution. 
\subsection{Asymptotic structure of the effective pressure}
As before, the $\mathcal{E}=0$ approximate solution \eqref{eq:NE0K2} governs $N$ for intermediate $x$, provided $K$ is not sufficiently large for this solution to break down (as is the case for the values chosen above). Much of this region corresponds to the Coulomb-dominated region, where the ice surface slope is very small, and $N$ is correspondingly small. 
(In this region the ice is close to flotation but, unlike in the extensional stress boundary layer, the basal friction and the driving stress balance while the extensional stress remains negligible.)
It can be seen in Figure \ref{fig:SS_varyK_m8}(c) that the small-$x$ and transition region solutions for $N$ break distinctly from this behaviour, since $N$ increases with increasing $K$ near the origin and the local maximum, rather than decreasing as predicted by the $\mathcal{E}=0$ approximate solution. 
To explain why this is the case, we  undertake an asymptotic analysis of these regions, beginning with the transition region.  

% \begin{equation}\label{eq:NE02}
% N\sim \frac{-\frac{\partial}{\partial x}{} \left( (1-\delta)H + b \right)}{mx + KH_{sb}  \frac{\partial}{\partial x}{} \left( (1-\delta)H + b \right)}, \quad \text{for} \quad x\gg \sqrt{\frac{\mathcal{E}}{m}}, \quad x_g-x\gg \frac{\mathcal{E}}{mx_g}
% \end{equation}
% * It would be nice to reckon with the Coulomb interior  solution if we have time to progress with it.

% * Why does the transition region grow so much?
From Figure \ref{fig:SS_varyK_m8}(e), it can be seen that the ice sheet approximately obeys the Weertman sliding law throughout the effective pressure transition region. 
The transition region is therefore governed by the generalisation of  Equations \eqref{eq:transH} and \eqref{eq:transN},
\begin{equation}\label{eq:transHK}
 \frac{\partial \hat{H}}{\partial \xi}  =\frac1{\hat{H}^{m_W+1}}  ,
\end{equation}
\begin{equation}\label{eq:transNK}
 \frac{\partial \hat{N}}{\partial \xi}  =\frac1{\hat{H}^{m_W+1}}    -  \frac{\hat{N}}{1+\hat{\kappa} \hat{N}},
\end{equation}
where
\begin{equation}
\hat{\kappa} =(1-\delta)\left(\frac{a^{m_W} m {x_g}^{m_W-1}}{\mathcal{E}^{m_W+1}} \right)^{1/(m_W+2)} KH_{sb}(x_g).
\end{equation}
Increasing $K$ for fixed $m$ corresponds to increasing $\hat\kappa$ (although the scalings also depend indirectly on $K$ via $x_g$). 
The generalisation of the matching condition \eqref{eq:match_trans} is given by
\begin{equation}\label{eq:match_transK}
\hat{N} \sim \frac{1}{\hat{H}^{m_W+1} - \hat{\kappa} } \quad \text{as} \quad \xi\to\infty.
\end{equation}

\begin{figure}
    \centering
    \includegraphics[width=.75\linewidth]{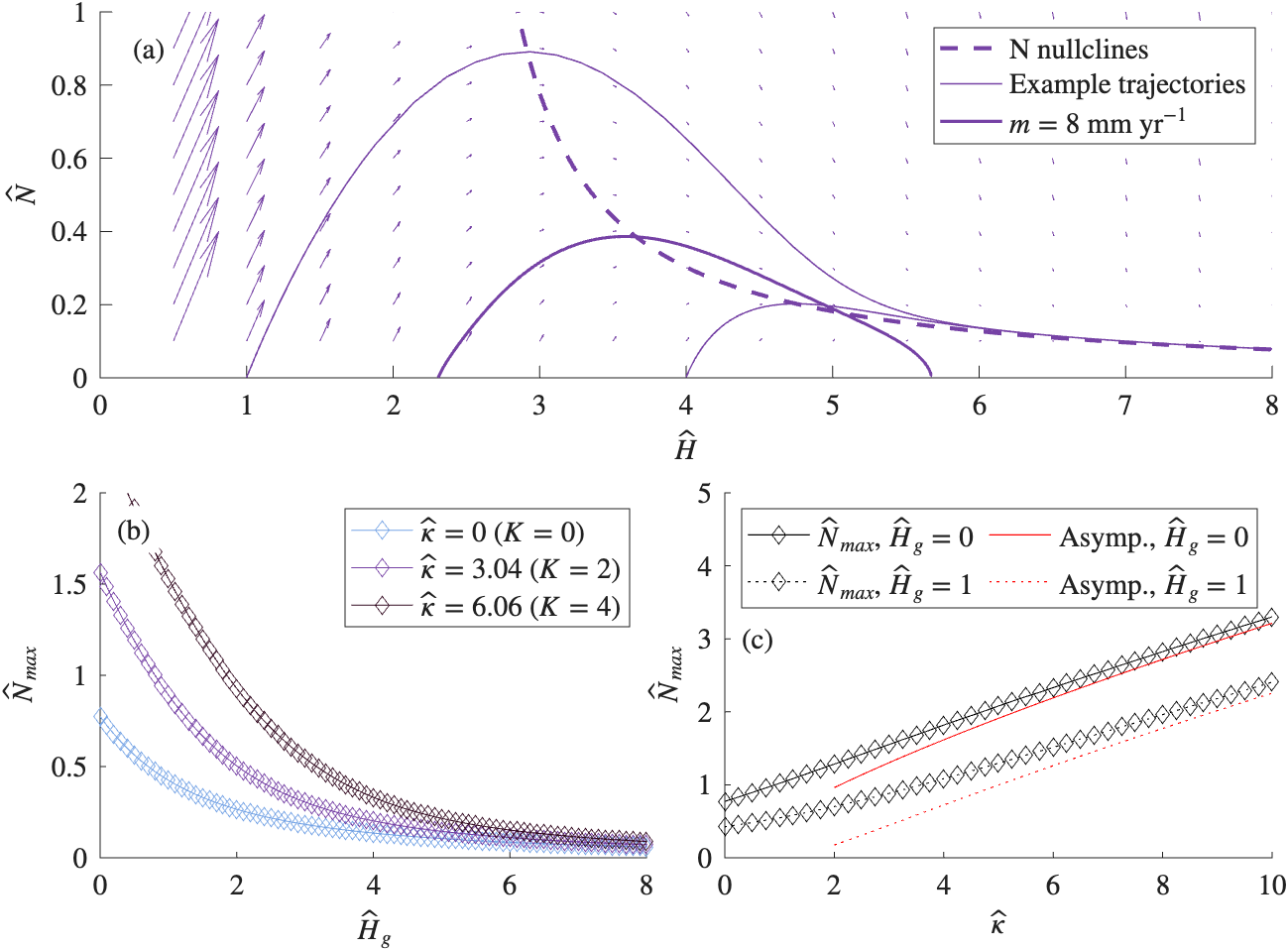}
    \caption{(a) Phase portrait of Equations \eqref{eq:transHK} and \eqref{eq:transNK} for selected $\hat \kappa=3.06$, corresponding to $K=2$ with $m=8$, including the scaled full solution as in Figure \ref{fig:trans_K0}(a). (b) Maximum $\hat{N}$ as a function of the initial condition $\hat{H}_g$, as in Figure \ref{fig:trans_K0}(b), for the same $\hat \kappa$ as in (a). (c) Maximum $\hat{N}$ as a function of $\hat{\kappa}$, for initial conditions $\hat{H}_g=0$ and $\hat{H}_g=1$, compared to the asymptotic solution \eqref{eq:Nmax_largeK}.}
    \label{fig:trans_Knon0}
\end{figure}

%Figure \ref{fig:trans_Knon0} shows the phase portrait of these Equations for various $\hat \kappa\geq0$, indicating the trajectory directions, the $\hat N$-nullcline, some example trajectories, and the scaled full solution in each case. 
Figure \ref{fig:trans_Knon0}(a) shows the phase portrait of these Equations for $m=8$ mm yr$^{-1}$ and $\hat \kappa = 3.04$ ($K=2$), indicating the trajectory directions, the $\hat N$-nullcline, some example trajectories, and the scaled full solution. 
(The latter is truncated in order to focus on where the transition region regime applies.) 
As with $\hat \kappa = 0$ (Figure \ref{fig:trans_K0}), the trajectories begin at $\hat{H}=\hat{H}_g$, $\hat{N}=0$, where $\hat{H}_g$ is the scaled flotation thickness defined by Equation \eqref{eq:Hg}. 
The trajectories initially rise before crossing the $\hat N$-nullcline at a maximum $\hat{N}_{max}(\hat{H}_g)$ and ultimately converging towards this nullcline from above, ensuring matching to the $\mathcal{E}=0$ approximate solution. 

Since the nullcline is given by
\begin{equation}
\hat{N} = \frac{1}{\hat{H}^{m_W+1}-\hat \kappa},
\end{equation}
increasing $\hat \kappa$ shifts this nullcline in the direction of increasing $\hat N$ and $\hat H$, with an asymptote at $\hat{H} = {\hat{\kappa}}^{1/(m_W+1)}$. 
As a result, the value of $\hat{N}_{max}$ attained for a given initial condition $\hat{H}_g$ is higher for larger $\hat{\kappa}$. Figure \ref{fig:trans_Knon0}(b) shows the shift of the function $\hat{N}_{max}(\hat{H}_g)$ as $\hat \kappa$ is increased. (Note that however that varying $\hat{H}_g$ for a fixed $\hat{\kappa}>0$ is difficult in practice, since varying $m$ also affects $\hat{\kappa}$.) The increased maximum also means that the solution takes longer to decay towards the limiting nullcline. After rescaling, this explains why the transition region of the effective pressure has a higher (local) maximum $N$ and is larger overall when $K$ is increased. 
As $\hat \kappa \to \infty$, Equations \eqref{eq:transHK} and \eqref{eq:transNK} are approximately
\begin{equation}
\frac{\partial \hat{H}}{\partial \xi} =  \frac1{\hat{H}^{m_W+1}} , \quad 
 \frac{\partial \hat{N}}{\partial \xi}  =\frac1{\hat{H}^{m_W+1}}    -  \frac{1}{\hat{\kappa}},
\end{equation}
which have the first integral
\begin{equation}\label{eq:Ntrans_largeK}
\hat N = \hat{H}-\hat{H}_g-\frac{\hat{H}^{m_W+2}-\hat{H}_g^{m_W+2}}{(m_W+2)\hat\kappa}.
\end{equation}
Finding the maximum of Equation \eqref{eq:Ntrans_largeK} gives the behaviour of $\hat{N}_{max}(H_g)$ for large $\hat{\kappa}$ as
%The maximum value of $\hat N$ in this limit is therefore given by the maximum
\begin{equation}\label{eq:Nmax_largeK}
{\hat N}_{max} = \left( \frac{m_W+1}{m_W+2}\right) \hat{\kappa}^{1/(m_W+1)} -\hat{H}_g+\frac{\hat{H}_g^{m_W+2}}{(m_W+2)\hat\kappa}.
\end{equation}
This behaviour is plotted against the value of $\hat{N}_{max}$ for various $\hat{\kappa}$ in Figure \ref{fig:trans_Knon0}(c). 
Taking the leading behaviour ${\hat N}_{max} \sim ((m_W+1)/(m_W+2))\hat{\kappa}^{1/(m_W+1)} $gives the re-scaled behaviour for large $K$
\begin{multline}
N_{max} =(1-\delta)\left(\frac{a^{m_W} {x_g}^{m_W-1}}{m \mathcal{E}^{m_W+1}} \right)^{1/(m_W+2)} \hat{N}_{max} \\\sim \left(\frac{m_W+1}{m_W+2} \right)\left(\frac{(1-\delta)^{m_W+2}a^{m_W} {x_g}^{m_W-1}  KH_{sb}(x_g)}{m \mathcal{E}^{m_W+1}} \right)^{1/(m_W+1)},
\end{multline}
showing that the local maximum of $N$ grows roughly with $K^{1/{(m_W+1)}}$ as $K$ increases.

% * Leave out this para?
% For larger values of $K$, the assumptions of this transition region  may break down before the trajectory reaches the limiting nullcline. 
% This is seen for the scaled full solutions in Figure \ref{fig:trans_Knon0}(a), where $N$ becomes low enough that the Weertman sliding law no longer applies, so that a separate theory is required to match to the $\mathcal{E}=0$ region. 
% For the case where $m=1$ mm yr$^{-1}$ (Figure \ref{fig:SS_varyK_m1}), the transition region encompasses the whole domain, and so the limiting $\mathcal{E}=0$ solution is never reached. 

% * We can also note the behaviour of the near 0 solution 
We can also investigate the effect of the sedimentary basin on the behaviour of $N$ for small $x$. 
For nonzero $K$, the equivalent of Equation \eqref{eq:NSS_near0} is
\begin{equation}\label{eq:NSS_near0K}
- \frac{(ax)^{m_W}}{H^{m_W+1}} - \mathcal{E} \frac{\partial N}{\partial x}   + \frac{mx N}{1+KH_{sb}N} =0.
\end{equation}
If we rescale the variables as 
\begin{equation}\label{eq:scalings_smallx}
x = \sqrt{\frac{\mathcal{E}}{m}} \tilde{x}, \quad N = \frac{a^{m_W} \mathcal{E}^{(m_W-1)/2}  }{H(0)^{m_W+1} m^{(m_W+1)/2} } \tilde{N},
\end{equation}
we obtain at leading order a single equation for $\tilde{N}$, decoupled from $\tilde{H}$,
\begin{equation}\label{eq:near0_scaled}
\frac{\partial \tilde N}{\partial \tilde x} = - \tilde{x}^{m_W}     + \frac{\tilde{x} \tilde{N}}{1 + \tilde{\kappa} \tilde{N}} ,
\end{equation}
where
\begin{equation}
\tilde{\kappa} = \frac{a^{m_W} \mathcal{E}^{(m_W-1)/2} }{H(0)^{m_W+1} m^{(m_W+1)/2} } K H_{sb}(0),
\end{equation}
since $H=H(0)$ and $H_{sb}=H_{sb}(0)$ are constant to leading order. 
Matching with the $\mathcal{E}=0$ approximate solution \eqref{eq:NE0} requires that
\begin{equation}\label{eq:asymp_near0}
\tilde N  \sim \frac{\tilde{x}^{m_W-1}}{1 - \tilde{\kappa} \tilde{x}^{m_W-1}} \quad \text{as} \quad \tilde{x} \to \infty.
\end{equation}
The solution of Equation \eqref{eq:near0_scaled} subject to Equation \eqref{eq:asymp_near0} gives an approximate solution for $N$ for $x = O(\sqrt{\mathcal{E}/m})$. 
In particular, we can derive the approximate behaviour of $\tilde N$ for $\tilde{\kappa}\gg 1$. 
In this case, so long as $\tilde N \gg \tilde{\kappa}^{-1}$, we have 
\begin{equation}\label{eq:smallx_bigN}
\frac{\partial \tilde N}{\partial \tilde x} \approx - \tilde{x}^{m_W}     + \frac{\tilde{x} }{\tilde{\kappa} } .
\end{equation}
This implies that $\tilde{N}$ decreases from an initial value $\tilde{N}(0)$ until reaching a local minimum at $x = \tilde{\kappa}^{1/(1-m_W)}$, whereafter $\tilde N$ increases. 
If $\tilde{N} \gg \tilde{\kappa}^{-1}$ for all $\tilde x$, then Equation \eqref{eq:smallx_bigN} implies that $\tilde N$ grows without bound as $\tilde x \to \infty$, which violates the matching condition \eqref{eq:asymp_near0}. 
Therefore, $\tilde{N}$ in Equation \eqref{eq:smallx_bigN} must reach $\tilde{\kappa}^{-1}$, and in particular its minimum value must satisfy
\begin{equation}
\tilde{N}\Big(\tilde{\kappa}^{1/(1-m_W)}\Big) \approx \tilde{N}(0)- \int_0^{\tilde{\kappa}^{1/(1-m_W)}}\Bigg( \tilde{x}^{m_W}     - \frac{\tilde{x} }{\tilde{\kappa} } \Bigg)\,\mathrm{d}\tilde{x}' = O\big( {\tilde \kappa}^{-1}\big).
\end{equation}
Consequently, we have 
\begin{equation}
\tilde N(0) \sim \int_0^{\tilde{\kappa}^{1/(1-m_W)}}\Bigg( \tilde{x}^{m_W}     - \frac{\tilde{x} }{\tilde{\kappa} } \Bigg)\,\mathrm{d}\tilde{x}'= \left(\frac{1}{m_W+1}-\frac12\right)\tilde{\kappa}^\frac{1+m_W}{1-m_W},
\end{equation}
so that
\begin{equation}\label{eq:smallN_bigK_sol}
\tilde N \sim \left(\frac{1}{m_W+1}-\frac12\right)\tilde{\kappa}^\frac{1+m_W}{1-m_W} - \frac{\tilde{x}^{m_W+1}}{m_W+1}     + \frac{\tilde{x}^2 }{2\tilde{\kappa} } \quad \text{for} \quad  \tilde{x} \ll \tilde{\kappa}^{1/(1-m_W)}.
\end{equation}
The leading-order solution for $\tilde{x} \gg \tilde{\kappa}^{1/(1-m_W)}$ is given by the asymptotic behaviour \eqref{eq:asymp_near0}. 
Equation \eqref{eq:smallN_bigK_sol}, together with the scalings \eqref{eq:near0_scaled}. gives the behaviour of $N$ for $x= O(\sqrt{\mathcal{E}/m})$ provided $K$ is sufficiently large. 
For $m_W = 1/3$, the exponent $(1+m_W)/(1-m_W)=2$, implying that the effective pressure at  the origin $N(0)$ grows quadratically with the permeability $K$ to leading order. 

\subsection{Summary}
The results of this subsection  indicate that groundwater flow in sedimentary basins has the potential to significantly affect the steady-state thickness and velocity of an ice sheet. 
This is because the sedimentary basin can drain water away from certain regions of the subglacial hydrological system, locally increasing the effective pressure, and exfiltrate this water elsewhere to create regions of low effective pressure in a self-reinforcing feedback. 
The inclusion of a sedimentary basin also results in a slightly more advanced steady grounding line position. 
%The sedimentary basin also affects the steady grounding line position, albeit to a lesser extent. 
We infer that it is important to consider sedimentary basin groundwater flow in models for subglacial hydrology and basal friction, and that such groundwater flow could help to explain regions of locally high or low basal friction. 

It should be noted, however, that in reality the effect of groundwater flow is unlikely to be as extreme as in Figure \ref{fig:SS_varyK_m8}. 
As mentioned above, this is because we have focused on the highest likely values of $k_{sb}$ for illustrative purposes. 
Our model also neglects certain dynamic effects that would likely prevent the predicted regions of near-flat ice surface from being achieved in reality. 
Notably, ice streams are typically confined between regions of slower-flowing ice, which restrict sliding of the ice by exerting a lateral shear stress, something that our one-dimensional ice sheet model neglects. 
We discuss this and other limitations of working in a single horizontal dimension in Section 6. 

\subsection{Effects of sedimentary basin geometry}

\begin{figure}[t]
    \centering
    \includegraphics[width=0.7\linewidth]{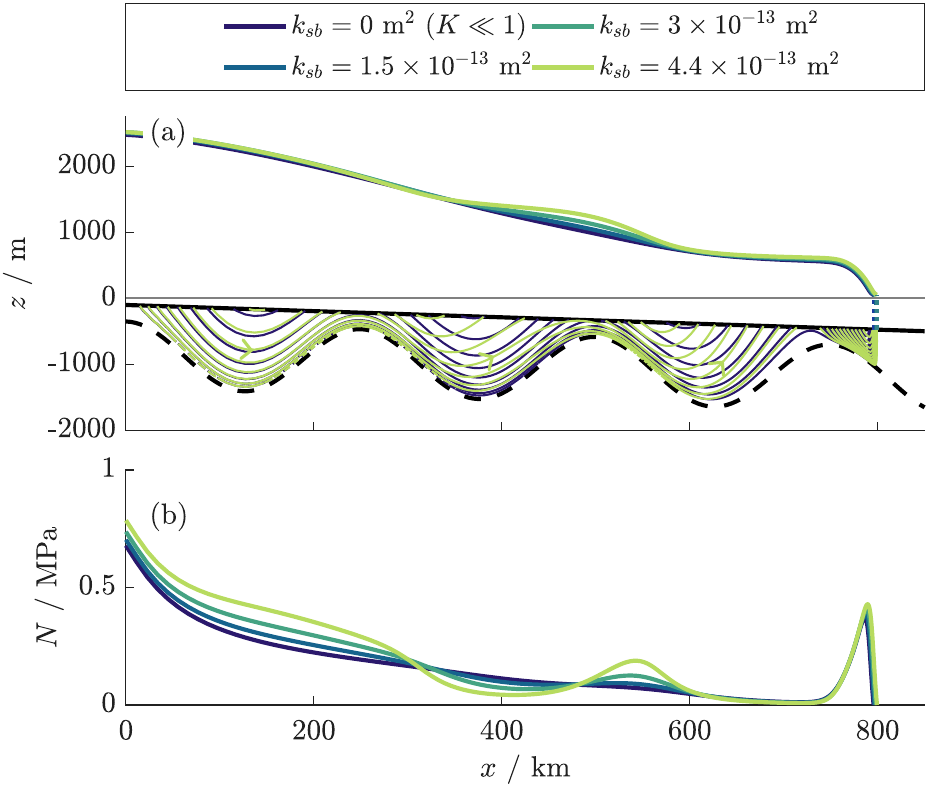}
    \caption{(a) Ice sheet profile and sedimentary basin streamlines and (b) effective pressure for a nonuniform sedimentary basin depth $H_{sb} = 750 - 500 \cos(2\pi x / 250 \text{ km})$ m, and for a melt rate of $m=8$ mm yr$^{-1}$ (dimensionless $m=8$). The various values of $k_{sb}$ correspond to $K\ll1$ and $K=1$, 2, 3, and the plotted streamlines are as in Figures \ref{fig:SS_varyK_m1} and  \ref{fig:SS_varyK_m8}. %, and otherwise as in Figure \ref{fig:SS_varyK_m8},  with various values of $K \geq0$. 
    (Note that in (a), as elsewhere, the aspect ratio is exaggerated for illustrative purposes.)}
    \label{fig:SS_nonu}
\end{figure}

So far we have worked with a uniform sedimentary basin thickness $H_{sb}$. However, previous work has found that shallowing and deepening of a sedimentary basin can drive exfiltration and infiltration respectively \citep{cairns2025groundwater}. 
We therefore investigate the consequences of variations in sedimentary basin geometry in Figure \ref{fig:SS_nonu}, using an idealised sedimentary basin thickness $H_{sb} = 750 - 500 \cos(2\pi x / 250 \text{ km})$ m. 
The ice sheet profile and streamlines for various $K$ are shown in Figure \ref{fig:SS_nonu}(a), and we can see from the latter that several regions of exfiltration and infiltration emerge. 

The effective pressure in Figure \ref{fig:SS_nonu}(b) shows the effect of this exfiltration and infiltration, which result in local peaks in $N$ where the sedimentary basin is deepening (infiltration) and troughs where the sedimentary basin is shallowing (exfiltration). 
These are most prominent in the middle of the domain, where the effective pressure follows the $\mathcal{E}=0$ approximate solution. 
The small-$x$ and transition region approximate solutions are independent of the sedimentary basin geometry to leading order, since this can be approximated as constant. Hence, in this region $N$ displays little influence due to sedimentary basin geometry.

In the ice sheet in Figure \ref{fig:SS_nonu}(a), we can see the effect of these changes in the effective pressure on the ice profile. 
The local trough in effective pressure produces a plateau-like region of flatter surface slope, whereas the local peak produces a steeper region, resulting in a staggered shape. 
The net effect of this steepening and flattening on the ice thickness further inland is negligible, so that the effect of the sedimentary basin geometry is localised. 

These results suggest, in addition to their permeability, the geometry of sedimentary basins is important in determining their effect on the ice sheet. %, in addition to their permeability. 
Moreover, they suggest that the ice sheet profile may give an indication of the properties of the underlying sedimentary basin. 
In particular, regions of steeper or flatter ice may suggest a change in basal friction, which could be explained by local variations in sedimentary basin geometry driving exfiltration and infiltration. 

\section{Groundwater feedback during grounding line motion}
% Effect of compaction-driven exfiltration for transient examples: advance/retreat to steady state. %, and periodic motion. 
% Fig x2: grounding line over time in each case, perhaps with `snapshots'

% Maybe a log plot would be useful in order to be qualitative.
\begin{figure}
    \centering
    \includegraphics[width=0.8\linewidth]{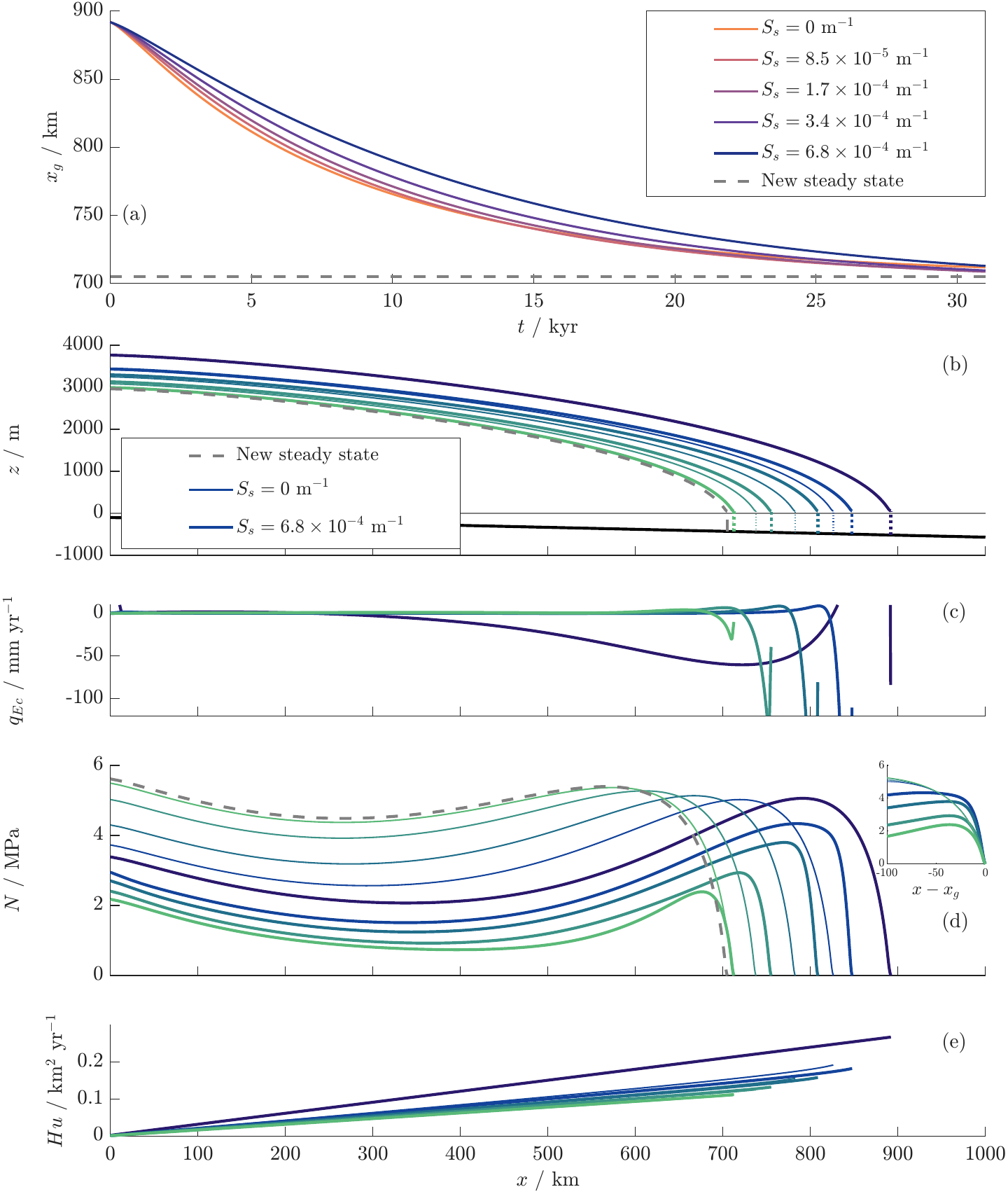}
    \caption{(a) Grounding line position $x_g(t)$, retreating to a steady state following a step decrease of the accumulation $a$, for various values of $S_s$ corresponding to $\Sigma = 0$,  60, 120, 240, 480 respectively, $m=1$ mm yr$^{-1}$, $k_{sb} = 1.6 \times 10^{-13}$ m$^2$, and otherwise as in Figure \ref{fig:SS_varyK_m1}. (b) Ice sheet thickness $H$, (c) compaction-driven exfiltration $q_{Ec}$ (see Equation \eqref{eq:qEc}), (d) effective pressure $N$, and (e) ice flux $H u$, at times $t=0$, 3.9, 7.8, 15.7 and 31 kyr, for $S_s=0$ and $6.8 \times 10^{-4}$ m$^{-1}$. The inset in panel (d) shows $N$ against $x_g-x$. A lighter and greener colour indicates later times.}
    \label{fig:retreat}
\end{figure}

We now turn our consideration to transient solutions of the model, in order to investigate the effect of  compaction-driven exfiltration (the second term in Equation \eqref{eq:qE}). Since this depends on $\partial H/\partial t$ and $\partial N / \partial t$, it does not appear in the previous steady state solutions, which only include topographically-driven exfiltration.
For this section we continue to include the latter (using a modest but nonzero $k_{sb}$), but focus on the effect of the former. 

A canonical transient solution is evolution of the system towards a steady state. 
This case can be brought about when a parameter of the model, such as the accumulation $a$, rapidly changes, meaning that the ice sheet is no longer in a steady state and must advance or retreat to a new one.   
Similar scenarios has been investigated in prior studies of ice and hydrology \citep[\textit{e.g.}][]{lu2023coupling}, and additionally resemble the behaviour of the ice after crossing `tipping points' that can emerge from non-monotone bed geometry \citep{schoof2007ice}. 
A change in the accumulation $a$ could act as a simple representation of climate change.
%A change in the temperature-sensitive $A$ could represent, for example, a change in the climate. 

In this section, we show that, when the ice is retreating towards a steady state, compaction-driven exfiltration provides a negative feedback that slows this retreat. 
Using the same linear prograde bed $b(x)$ as above, and with constant sedimentary basin thickness $H_{sb}=2000$, melt rate $m=1$ mm yr$^{-1}$ and $k_{sb}=1.6 \times 10^{-13}$ m$^2$, we initialise the system in a steady state with $a=0.3$ m yr$^{-1}$.  %$A = 1 \times 10^{-25}$ s$^{-1}$ Pa$^{-3}$. 
%We then step-change the value at time $t=0$ to $A = 2 \times 10^{-25}$ s$^{-1}$ Pa$^{-3}$, so that the steady state is moved further back, requiring the ice sheet to retreat towards the new steady state. 
We then step-change the value to $a=0.15$ m yr$^{-1}$, so that the steady-state grounding line is moved further back, requiring the ice sheet to retreat towards the new steady state. %The accumulation follows a function
% \begin{equation}
% a(t) = (0.15 + 0.15 \exp(- t / 3900 \, \mathrm{yr})) \, \mathrm{m \, yr}^{-1}.
% \end{equation}
We repeat this scenario for different values of $\Sigma$, the parameter determining the strength of compaction-driven exfiltration, which are achieved by varying the specific storage $S_s$. 
As mentioned previously, realistic values of $S_s$ give $\Sigma$ ranging from 0.07 to 70, but in this section we include some larger-than-realistic values of $\Sigma$ for illustrative effect. 

The results of this experiment are shown in Figure \ref{fig:retreat}. 
Figure \ref{fig:retreat}(a) compares the grounding line position $x_g(t)$ over time for various values of $\Sigma$. 
We can see that increasing $\Sigma$ (via $S_s$) makes this retreat slower. %, with the effect beginning early on and persisting throughout the experiment. 
The largest difference is seen between around 5--15 kyr, roughly following where  the curve $x_g(t)$ is steepest. % without compaction-driven exfiltration. 
%However, towards the end of the experiment, the cases including compaction-driven exfiltration catch up to the case without, particularly for intermediate values of $S_s$. 
The mechanism behind this acceleration is shown in Figures \ref{fig:retreat}(b), (c), (d) and (e) which show the ice sheet profile, the compaction-driven part of the exfiltration $q_{Ec}$ (see Equation \eqref{eq:qEc}), the effective pressure $N$, and the ice flux $H u$ at selected points in time, for the cases with and without compaction-driven exfiltration ($\Sigma=0$ and $\Sigma=480$). %$S_s=0 $ m$^{-1}$ and its maximum value ($\Sigma=0$ and $\Sigma=480$). 

% * EXPLANATION
As seen in Figure \ref{fig:retreat}(b), the retreat of the ice leads to ice thinning ($\partial H / \partial t<0$), which is most pronounced near the grounding line as the surface slope is steepest. 
Seen in Figure \ref{fig:retreat}(c), this leads to a small amount of exfiltration in the interior of the ice sheet, but a larger amount of infiltration in a small region close to the grounding line. 
%When compaction-driven exfiltration is included, this thinning ($\partial H / \partial t<0$) without an immediate corresponding change in $N$ produces an exfiltration $q_{Ec}$, seen in Figure \ref{fig:retreat}(c), which is strongest near the grounding line. 
As a result, the effective pressure $N$ (Figure \ref{fig:retreat}(d)) is lowered compared to when $\Sigma =0$ away from the grounding line (as exfiltration increases the effective water thickness $h$), but increases compared to the $\Sigma =0$ case within the region of infiltration near the grounding line. 
The relative increase near the grounding line can be seen more clearly in the inset panel (noting that line colours correspond to the same time rather than the same $x_g$). 
The local increase in $N$ increases the basal friction near the grounding line, which decreases the ice flux across the grounding line (Figure \ref{fig:retreat}(e)), and hence slows the rate at which the ice retreats. 
Note that at the final timestamp depicted in Figure \ref{fig:retreat}, for $\Sigma=480$, the ice thickness and grounding line position are close to the steady state, but the effective pressure is still far from this state. 
This is because exfiltration away from the grounding line lowers the effective pressure in the interior region without influencing the grounding line dynamics.

% Although this lowering of $N$ occurs across the domain, its effect on the ice sheet is largest near the grounding line. 
% This exfiltration lowers the effective pressure, seen in Figure \ref{fig:retreat}(c), by increasing the effective water thickness $h$. 
% Although this lowering of $N$ occurs across the domain, its effect on the ice sheet is largest near the grounding line. 
% where it occurs via the first-order correction to the $\psi=0$ approximate solution. 
%Here, the reduction in basal friction leads to faster sliding of the ice and hence a higher ice flux across the grounding line, resulting in a faster rate of grounding line retreat. 
% This results in reduced basal friction near the grounding line, which leads to faster sliding of the ice and hence a higher flux of ice across the grounding line, producing accelerated grounding line retreat. 

To explain why compaction drives exfiltration over most of the domain but infiltration near the grounding line, we revisit the governing equation \eqref{eq:hyd*2}. 
Assuming that $\lambda \ll \Sigma \mathcal{E}$, as is the case for the relatively large $\Sigma$ considered here, this equation can be approximately rearranged to
\begin{multline}\label{eq:hyd_lam0}
0 =   \frac{\partial}{\partial x}{} \left[ \left(h + K H_{sb}\right)    \frac{\partial}{\partial x}{} \left( (1-\delta)  H- \mathcal{E} N +  b\right)  \right] + m \\- \Sigma H_{sb} \left(  (1-\delta)(1-\xi)\frac{\partial H}{\partial t}- \mathcal{E}   \frac{\partial N}{\partial t} \right)  .
%- \Sigma \mathcal{E}  H_{sb}  \frac{\partial N}{\partial t} =   \frac{\partial}{\partial x}{} \left[ \left(h + K H_{sb}\right)    \frac{\partial}{\partial x}{} \left( (1-\delta)  H- \mathcal{E} N +  b\right)  \right] + m \\- \Sigma H_{sb}   (1-\delta)(1-\xi)\frac{\partial H}{\partial t} .
\end{multline}
Taking $\lambda\approx 0$ means that $h$, and hence $N$, responds instantaneously to the current values of the source terms (melting and exfiltration/infiltration). 
However, the compaction-driven exfiltration itself depends on the resulting rate of change of $N$. 

Away from the grounding line, $\mathcal{E}N$ is smaller than $(1-\xi)H$. 
This means that ice thinning ($\partial H/ \partial t <0$) leads to a net exfiltration, as this term dominates the compaction-driven exfiltration (via the expansion of the pore water). 
This exfiltration increases $h$, leading to a decrease in the effective pressure ($\partial N/ \partial t <0$). 
However, crucially, the rate at which $N$ decreases is not large enough to change the overall sign of the compaction term. 

% as the decrease in $N$ ($\partial N/ \partial t <0$) as $h$ increases due to this exfiltration is not large enough
%Away from the grounding line, $\mathcal{E}N$ is smaller than $(1-\xi)H$, and this equation has the effect that ice thinning ($\partial H/ \partial t <0$) lowers the effective pressure $N$. 
%This is because the compaction-driven exfiltration is dominated by the effect of $\partial H/ \partial t$ on $\partial p_w / \partial t$, which drives water expansion when the ice is thinning. 
%This can be seen in the reduction of $N$ away from the grounding line when $\Sigma>0$ in Figure \ref{fig:retreat}(d). 

However, close to the grounding line, as in the steady state, $N$ follows the $\psi=0$ approximate solution \eqref{eq:Npsi0} as a result of the boundary condition $N=0$ at $x=x_g$. 
This implies that when $H$ decreases, $N$ must decrease accordingly,
\begin{equation}
\frac{\partial N}{\partial t} \approx \frac{1}{\mathcal{E}} \frac{\partial H}{\partial t} \quad \text{near} \quad x=x_g.
\end{equation}
The compaction term is therefore approximately
\begin{equation}
q_{Ec} = - \Sigma H_{sb} \left(  (1-\delta)(1-\xi)\frac{\partial H}{\partial t}- \mathcal{E}   \frac{\partial N}{\partial t} \right) 
\approx \Sigma H_{sb} (1-\delta)\xi \frac{\partial H}{\partial t},
\end{equation}
so that ice thinning ($\partial H/ \partial t <0$) instead leads to a net infiltration. 
This infiltration is driven by dilation of the porous medium, while the expansion of the pore water is negligible because the pore pressure remains close to hydrostatic. 
Because this infiltration exceeds the total exfiltration over the interior (\textit{i.e.} $\int_0^{x_g}q_{Ec} \, \mathrm{d}x <0$), the resulting effect is that $N$ is higher near the grounding line for $\Sigma>0$ than for $\Sigma =0$. 

% This infiltration causes $N$ to be higher near the grounding line when compaction is included. Specifically, if we write 
% \begin{equation}
% N = \frac{1}{\mathcal{E}}\left( (1-\delta)H+ b \right) + \tilde{N},
% \end{equation}
% where $\tilde{N}$ is a small perturbation, we find that 
% \begin{equation}
%   - \Sigma  H_{sb} \frac{\partial \tilde{N}}{\partial t}   =  \frac{\partial}{\partial x} \left[ - \left(h + K H_{sb}\right) \mathcal{E}   \frac{\partial \tilde{N}}{\partial x}  \right] + m + \Sigma H_{sb}   (1-\delta)\xi\frac{\partial H}{\partial t} ,
% \end{equation}
% so that ice thinning ($\partial H / \partial t <0$) increases $\tilde{N}$ near the grounding line (though overall we still have $\partial N / \partial t <0$ due to the decrease in $H$). 
% As a result, $N$ is higher near the grounding line for $\Sigma>0$ than for $\Sigma =0$. (The perturbation $\tilde{N}=-\psi/\mathcal{E}$ is proportional to the hydraulic potential.)

\begin{figure}[t]
    \centering
    \includegraphics[width=.8\linewidth]{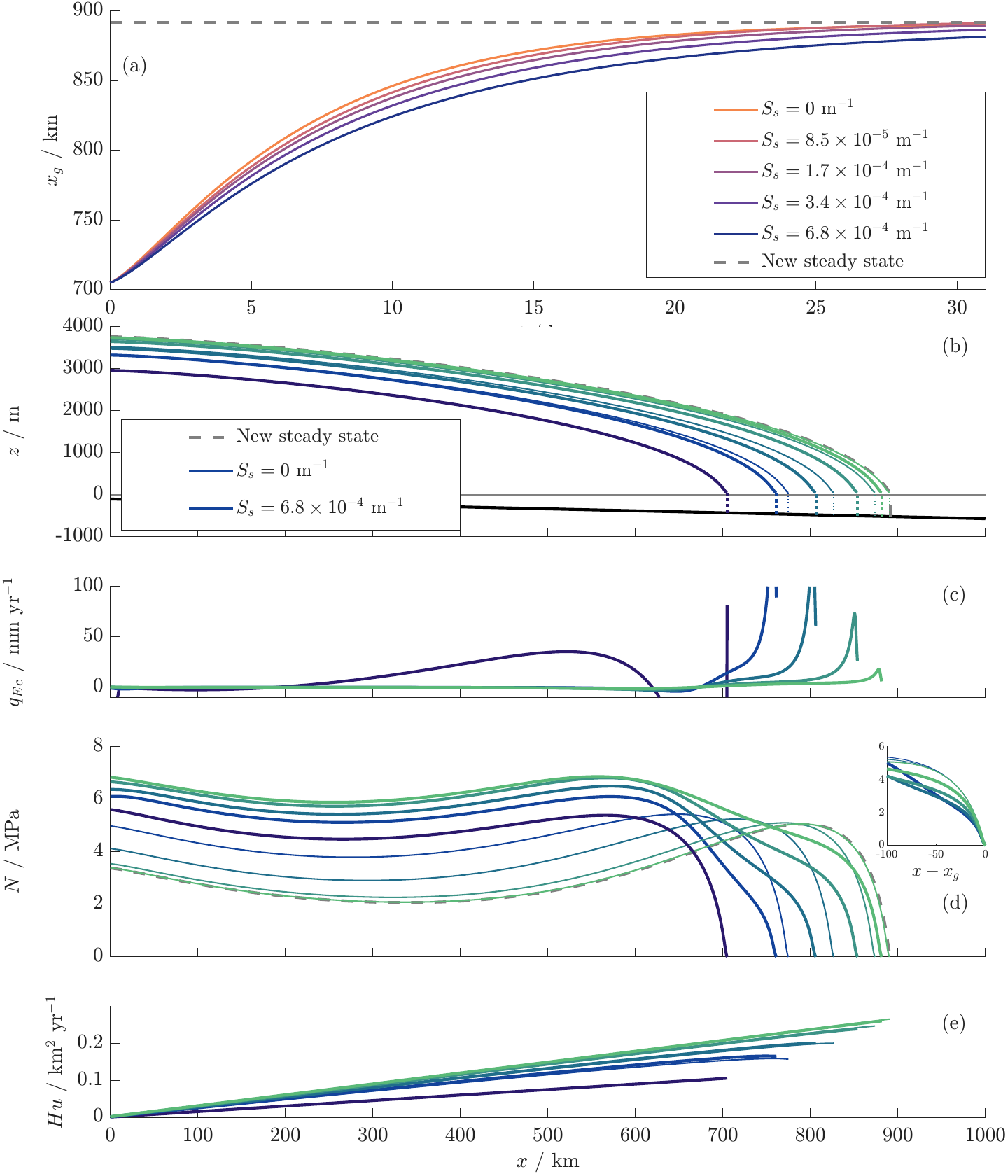}
    \caption{(a) Grounding line position $x_g(t)$, advancing to a steady state following a step increase in accumulation from $a=0.15 \, \mathrm{m \, yr}^{-1}$ to 0.3 m yr$^{-1}$ at time 0, and otherwise as in Figure \ref{fig:retreat}. (b) Ice sheet thickness $H$, (c) compaction-driven exfiltration $q_{Ec}$, and (d) effective pressure $N$, as in Figure \ref{fig:retreat}.}
    \label{fig:advance}
\end{figure}

We can also consider grounding line advance (rather than retreat) following a step increase in $a$, a case of which is shown in Figure \ref{fig:advance}. 
In this case, the ice thickens rather than thins, so that compaction drives infiltration and increased $N$ away from the grounding line, but drives exfiltration and reduced $N$ near the grounding line (again, as seen in the inset of panel (d)). 
The reduced effective pressure lowers the basal friction, increasing the ice flux across the grounding line. 
This leads to slower advance of the ice, because ice flows into the ice shelf rather than accumulating behind the grounding line. 
% Figure \ref{fig:advance} provides a counterpart to Figure \ref{fig:retreat} for the case of advance to confirm this behaviour. 

% Figure \ref{fig:advance} is the counterpart of Figure \ref{fig:retreat} for the case where $a$ instead undergoes a step increase, leading to grounding line advance. 
% We can confirm from Figure \ref{fig:advance} our hypothesis that compaction drives infiltration and increases $N$ far from the grounding line, but near the grounding line drives exfiltration and lowers $N$. 
% This reduces the basal friction and hence increases the ice flux across the grounding line, leading to slower advance of the grounding line. 

%This infiltration will increase the effective pressure and hence the basal friction, which results in slightly faster advance of the grounding line. 
%However, the effect is weaker than its counterpart during retreat, because the grounding line moves away from where the effective pressure is most affected rather than towards it. 

\begin{figure}
    \centering
    \includegraphics[width=1\linewidth]{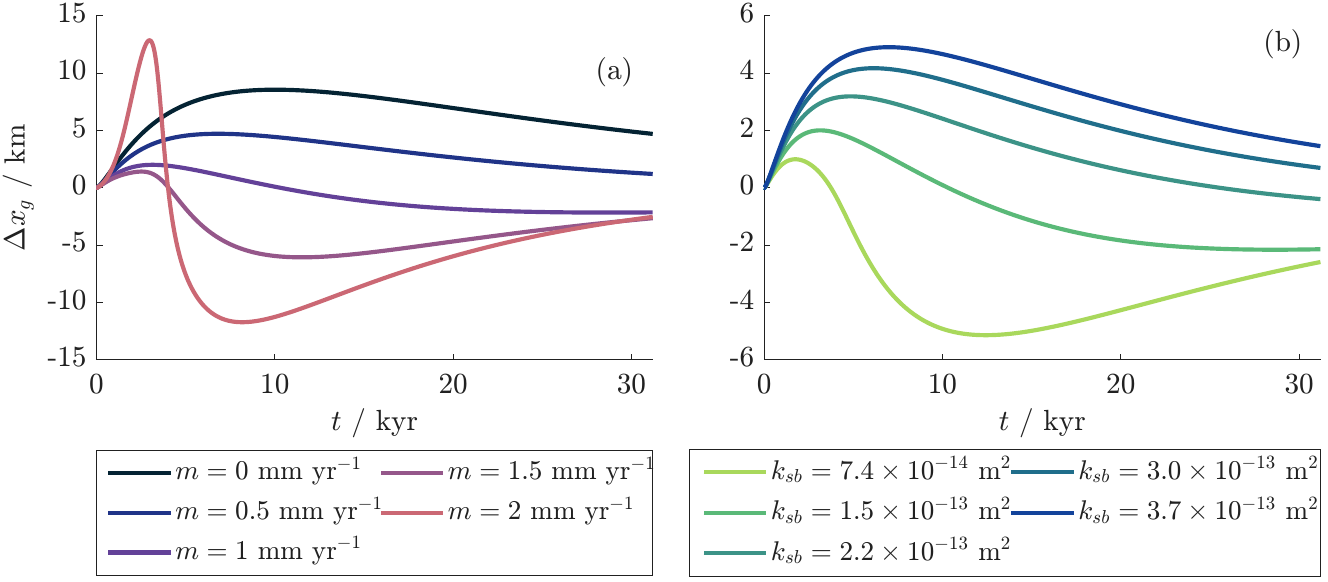}
    \caption{Difference in grounding line position $\Delta{x_g(t)} = x_g(t;\Sigma=30)-x_g(t;\Sigma=0)$ between the cases when $S_s=4.3 \times 10^{-5}$ m$^{-1}$   and $S_s=0$ m$^{-1}$ ($\Sigma=30$ and $\Sigma = 0$) for the retreat scenario considered in Figure \ref{fig:retreat}. We expect $\Delta{x_g(t)}>0$ if the retreat is slowed, as in Figure \ref{fig:retreat}. This is plotted for (a) varying $m$ corresponding to dimensionless $m=0$, 0.5, 1, 1.5, 2 with fixed $k_{sb}=1.6 \times 10^{-13}$ m$^2$ (b) varying $k_{sb}$ corresponding to $K=0.5$, 1, 1.5, 2, 2.5 with fixed $m=1$ mm yr$^{-1}$.}
    \label{fig:retreat_mK}
\end{figure}

A remaining question is how the strength of the response to compaction-driven exfiltration and infiltration varies with $m$ and $K$. 
We investigate this in Figure \ref{fig:retreat_mK}, by plotting $\Delta x_g(t) = x_g(t;\Sigma=30)-x_g(t;\Sigma=0)$, the relative difference in the grounding line positions between the cases when $S_s=4.3 \times 10^{-5}$ m$^{-1}$  and $S_s=0$ m$^{-1}$ ($\Sigma=30$ and $\Sigma = 0$), over time for a range of $m$ (Figure \ref{fig:retreat_mK}(a)) and $k_{sb}$ (Figure \ref{fig:retreat_mK}(b)). A positive value of $\Delta x_g$ implies that retreat is slowed. 
Generally, the effect is stronger if less water is present at the bed overall, since compaction-driven exfiltration then has a comparatively larger effect on the basal hydrology. This means the feedback is stronger if $m$ is lower or if $K$ is higher. However, for higher values of $m$, or lower values of $K$, the effect is weakened or even reversed, so that compaction accelerates retreat rather than slowing it.  
We hypothesise that this is because when the effective pressure is lower overall, the compaction-driven exfiltration and lowering of $N$ occurring beneath the interior of the ice sheet extends near enough to $x_g$ to influence the extensional stress boundary layer of the ice sheet. 
Once $m$ is sufficiently high (or $K$ sufficiently low), the overall exfiltration exceeds the infiltration  (\textit{i.e.} $\int_0^{x_g}q_{Ec} \, \mathrm{d}x >0$), so that $N$ is lowered overall at the grounding line rather than increased.

The results of this section indicate that, as well as affecting the steady states, sedimentary basins can also influence the transient dynamics of the ice-sheet-hydrology system through the separate process of compaction-driven exfiltration and infiltration. 
In particular, these process provide a feedback that slows both retreat and advance of the grounding line, which has potential significance when projecting the future of Antarctic ice streams that overlie sedimentary basins and are currently undergoing grounding line motion. 
These results depend on the  behaviour of the effective pressure near the grounding line, and are sensitive to the melt rate and sedimentary basin permeability as a result.  

\section{Discussion}

In this paper, we have developed an idealised two-way coupled model of ice flow and subglacial hydrology, including a sedimentary basin which exchanges groundwater with the basal hydrological system. 
For a steady state, we found that exfiltration generally occurs close to the grounding line and infiltration close to the ice divide, but that exfiltration can also occur inland if the ice profile is becoming flatter in the direction of flow. %, with the transition controlled by the second derivative of the hydraulic potential. 
Exfiltration and infiltration can also be driven by shallowing and deepening of the sedimentary basin respectively. 

Due to their effect on the water balance in the subglacial drainage system,  regions of exfiltration locally lower the effective pressure, and infiltration increases the effective pressure. 
If the sliding of the ice is dominated by the subglacial effective pressure, then these changes in effective pressure in turn lead to a steeper ice profile above areas of infiltration and a shallower plateau-like profile above areas of exfiltration. 
Overall, a deeper or more permeable sedimentary basin leads to thicker and slower-flowing ice in the steady state, due to certain localised increases in effective pressure. 

Transient changes in the basal water pressure and effective pressure also produce compaction-driven exfiltration and infiltration. 
During ice sheet retreat (for example due to a changing climate), compaction-driven infiltration near the grounding line can decelerate the process of retreat by increasing the effective pressure, and hence the basal friction, near the grounding line. 
Conversely, compaction-driven exfiltration near the grounding line can decelerate grounding line advance by reducing the effective pressure. 
% Transient changes in the basal water pressure also produce compaction-driven exfiltration when the water pressure decreases, and infiltration when it increases. 
% During ice sheet retreat (for example due to a changing climate), such exfiltration accelerates the process of retreat by lowering the effective pressure, and consequently basal friction, near the grounding line. 
% Conversely, infiltration during ice sheet advance accelerates this advance, but does so to a much lesser extent. 

Our model includes a number of idealising assumptions, in order to be simple to use and easy to interpret. 
However, there are a number of ways in which the theory in this paper could be extended to represent other features of Antarctic groundwater flow. 

Firstly, the model considers only one horizontal dimension, limiting its ability to be applied in modelling specific Antarctic ice streams. 
%For instance, the equations of mass conservation are unable to capture tributary sources of ice or basal water, or lateral variations in sedimentary basin geometry.
Lateral variations in topography could influence both ice flow and the routing of subglacial water. 
% Moreover, topographic variation could lead basal water to be routed in several different ways %depending on the topography, 
% rather than simply flowing towards the grounding line as assumed in this model.
Moreover, laterally-confining shear stresses could play a significant role on ice stream dynamics, and in particular could explain why regions of flat ice surfaces as seen in this paper are seldom observed in practice. 

In addition, lateral narrowing or broadening of the sedimentary basin could drive groundwater exfiltration or infiltration respectively, similarly to vertical shallowing and deepening.  
Nonetheless, our definition of exfiltration and infiltration $q_E$ in Equation \eqref{eq:qE} from sedimentary basins in this paper can easily extended to two horizontal dimensions as 
\begin{equation}
q_E =  \nabla \cdot \left[ \frac{k_{sb}}{\eta} H_{sb} \nabla \left( \rho_i g H - N + \rho_w g b \right) \right] - \frac{S_s}{\rho_w g} H_{sb}  \left( (1- \xi) \rho_i g   \frac{\partial H}{\partial t}  - \frac{\partial  N}{\partial t} \right),
\end{equation} 
and could therefore be coupled into existing state-of-the-art models for subglacial hydrology and ice flow. 
% - mass conservation x3, - ISM has buttressing forces
% Another limitation is the use of only one horizontal dimension, which neglects effects such as buttressing forces on the ice shelf \citep{gudmundsson2013ice}. 
% - hydrology model
%Including a second horizontal dimension could have equally important for the hydrology model, as basal water may accumulate in topographic lows rather than simply flowing towards the grounding line. 
% Sedimentary basin geometry
% - sed basin
% In addition, lateral variations in the geometry  of a sedimentary basin may lead to similar variations in exfiltration, and hence effective pressure and ice thickness and velocity, as we have discussed in Section 2.3.4. 

% - Other properties of the sedimentary basin
Our model also assumes an isotropic and homogeneous permeability of the sedimentary basin, although this assumption is unlikely to be accurate in reality.  
The assumption of isotropy is in fact unnecessary, since our model uses the Dupuit approximation, which treats vertical groundwater flow as instantaneous, so that only the horizontal permeability matters. %the use of a depth-integrated model based on the Dupuit assumption treats vertical groundwater flow as instantaneous, meaning that only the horizontal permeability is important. 
The depth-integrated approach can also be extended to heterogeneous $k_{sb}$, by replacing $k_{sb}$ with a depth average \citep{cairns2025groundwater}. 
However, if the permeability is highly heterogeneous (\textit{e.g.} due to the presence of faults, or large changes between rock strata), a Dupuit model may be altogether inappropriate, with a full two-dimensional model of groundwater flow required instead. 
% Variations in the porosity of the sedimentary basin may also affect the volume of groundwater that can be stored in a vertical column of sediment. 
% To account for this, the sedimentary basin thickness $H_{sb}(x)$ could be replaced by an `effective thickness', equivalent to $H_{sb}$ scaled with the depth-average of the porosity.

% Bedrock
In this paper, we have only considered one choice of bed topography $b(x)$, given by a linear prograde bed. 
Previous studies have shown that more realistic bed topographies can lead to  complex ice dynamics, such as multiple steady states, hysteresis, and bifurcations as parameters are varied \citep{schoof2007ice}. 
Although such cases are beyond the scope of this paper, our findings concerning the steady states and dynamic behaviour of this model generalise to more complex geometries. 
For instance, we expect compaction-driven exfiltration and infiltration to decelerate the grounding line retreat or advance that occurs after a bifurcation emerging from bed topography is crossed. 

% obviously A more complicated hydrology system
We have also only considered a single sliding law, the regularised Coulomb law \eqref{eq:rc}, in coupling the ice sheet to the hydrological system. 
A more comprehensive study could consider a range of sliding laws, along with different values of parameters such as $\mathcal{C}$ and $m_W$ \citep[as in \textit{e.g.}][]{brondex2019sensitivity, gregov2023grounding}. 
However, the results of this study should generalise to any sliding law in which the basal stress is an increasing function of effective pressure, with the caveat that the parameters of the sliding law will determine the overall strength of the coupling. 

%A limitation of the model itself is that we consider only a simple cavity-sheet model of subglacial hydrology. 
Our treatment of subglacial hydrology, based on a simple cavity-sheet model, is also deliberately simplified in order to be easy to solve and understand. 
While a simple model of this form ought to broadly agree with more complex models over the long timescales we consider \citep{defleurian2018shmip}, it may not be appropriate if the drainage system is dominated by flow in subglacial channels. %overlooks many of the means by which basal water is transported, such as subglacial channels. 
%A model of this sort may therefore be inappropriate when considering shorter timescales (\textit{e.g.} months to years), over which channels play an important role in Antarctic subglacial hydrology \citep{dow2022role}. 
Channels would be expected to form if the subglacial water flux becomes sufficiently large, and lead to an instability where larger channels drain water away from smaller ones \citep{flowers2015modelling}. 
% * cf boulton 2007 subglacial drainage and rempel 2008 effective stress
Due to their dynamics, effective pressure is generally increased near  such channels and lowered between them \citep{boulton2007subglacial, rempel2009effective}.
%Due to this water routing, effective pressure is lowered beneath channels and elevated between them, so compaction of the sedimentary basin could work against this instability by driving infiltration beneath channels and exfiltration between them. 
However, the modelling of subglacial channels requires consideration of thermodynamic effects, since channels open by melting in response to heat dissipation by the subglacial flow, which we discuss further below.

Even within the cavity-sheet framework, we have made simplifying assumptions, such as the form of the relationship $h(N)$, and in particular the fact that $h \to \infty$ as $N \to 0$. 
As mentioned above, a more accurate model would have $N \to 0$ once $h$ exceeds a finite height corresponding to the depth at which topographic obstacles are completely submerged. 
The use of such a model may modify the `$\psi=0$' approximate solution for $N$ here (Equation \eqref{eq:Npsi0}), and consequent results such as the feedbacks on grounding line motion considered in Section 4.5, although the `$\psi=0$' approximate solution should remain roughly accurate provide the limiting value of $h$ is relatively large. 
% In this model, the fact that $h \to \infty$ at the grounding line underpins the `$\psi=0$' approximate solution for $N$ here (Equation \eqref{eq:Npsi0}). 
% Therefore, using a different relationship $h(N)$ may modify some results of this model, such as the groundwater feedback during retreat considered in Section 5, which relies on the water pressure being close to hydrostatic near the grounding line. 
These considerations highlight the need for future work incorporating groundwater flow into a wider range of subglacial hydrology models, in particular models that are able to account for different modes of subglacial water transport.
% Once again, this suggests that future work could focus on incorporating sedimentary basins into more detailed state-of-the-art models of subglacial hydrology, which are able to describe several modes of subglacial water transport.

For simplicity, we have treated the melt rate as spatially uniform and prescribed, for example by a uniform geothermal heat flux. 
This neglects melting due to frictional heating, which depends on the ice velocity and basal stress, introducing another coupling to the model. 
However, for fast-flowing ice streams, such heating is likely to be significant, and even to dominate the energy budget. 
For example, for a Weertman sliding law with our choice of $C_W$, a sliding velocity of 200 m yr$^{-1}$ implies a frictional heat flux of $\tau_b u = C_W {u}^{m_W+1}=$ 890 mW m$^{-2}$, compared to estimated geothermal heat fluxes ranging from 30--150 mW m$^{-2}$ across Antarctica \citep{burton2020geothermal, staal2021antarctic}. 

A more accurate modelling approach would calculate the melt rate $m$ using an energy balance at the bed. 
%For example, rather than being treated as a constant parameter, the melt rate $m$ could be calculated using an energy balance at the bed \citep[as in \textit{e.g.}][]{haseloff2025subglacial}. 
This energy balance should include geothermal heating and basal friction, as well as conduction into the ice. 
Calculated in this way, $m$ would increase with $x$ following the ice velocity over most of the domain, but decrease near the grounding line, where the friction vanishes. 
Such an energy balance would introduce further coupling between the ice and hydrology, since for rapid sliding the friction $\tau_b$ is approximately $C_CN$, yet a higher friction leads to a lower $N$ by increasing the melt rate $m$. 
A model for an ice stream coupled via sliding to subglacial hydrology that includes such an energy balance among other thermal effects has been considered by \citet{haseloff2025subglacial}. 
Including such an energy balance could allow for further investigation of thermal effects. 
For example, it has been proposed that groundwater flow in sedimentary basins could enhance the geothermal heat flux to the ice bed \citep{gooch2016potential}, which would enable the sedimentary basin to exert feedback on basal melting. 
%It would also enable the inclusion of basal freeze-on as an additional driver of exfiltration, as previously considered by \citet{christoffersen2003thermodynamics}.
% * Talk more about frozen fringe formation and Schoof's stuff.

Incorporating thermodynamics into the model could moreover allow for a description of basal freeze-on and frozen fringe formation, which would be expected for the higher effective pressures in this paper \citep{rempel2008theory}. 
Where the bed is frozen, the permeability of the subglacial hydrological system is effectively zero, and the sliding of the ice is also inhibited \citep{meyer2018freeze, hansen2024presence}. 
In such a region, the `exfiltration' must match the rate at which ice is accreted onto the frozen fringe \citep{christoffersen2003thermodynamics}. 
In reality, large parts of the Antarctic ice-bed interface are likely to be frozen, and a hydrology model incorporating this should include a frozen region as a free boundary. This free-boundary problem is important in enabling ice streaming, with ice streams above thawed bed confined by slower-moving ice above frozen bed \citep{schoof2021role}. 
A potential consequence of groundwater flow for this problem is that groundwater can flow beneath regions of frozen bed. 
This allows for the thawed and frozen regions to exchange water and heat with one another, and could moreover enable water transport between two thawed regions separated by a frozen region. 

Overall, we have demonstrated how groundwater exchange with sedimentary basins may be incorporated into a model of subglacial hydrology coupled to ice sheet flow. 
Our research gives new insights into how such groundwater exchange could influence subglacial hydrology and, consequently, the dynamics of the ice sheet. 
This work highlights the need for further modelling, combined with observational data, in order to better understand the role of sedimentary basins in Antarctic subglacial hydrology, and the implications for Antarctica's contribution to future sea level rise. 

\paragraph{Acknowledgments}
We thank Christian Schoof and two anonymous referees for reviewing this paper and providing  constructive feedback. 
We also thank Richard Katz and Dominic Vella for useful feedback on an early version of this manuscript.

\paragraph{Funding statement}
This research has been supported by UK Research and Innovation (grant no. EPSRC-2022).

\paragraph{Competing interests}
The authors declare none.

\paragraph{Data and code availability}
Code and data used to generate the figures in this paper may be found at \url{https://doi.org/10.5281/zenodo.23314554}.

\paragraph{Author contributions}
GJC, GPB, and IJH developed the model. GJC wrote the code and produced the results. GJC wrote the paper with input from GPB and IJH.

{\normalsize \bibliography{shallow_bib}}

\appendix

\section{Non-dimensionalisation}

% Do we need to add some sources here
\begin{table}
\centering
\begin{tabular}{ c| c| c | c}
Parameter     & Value Used (Range) & Units & Source  \\
\hline
$[x]$    & $5\times10^5$ & m & \\
$a$  & 0.3 (0.1--0.5)  & m yr$^{-1}$ & \citet{kaspari2004climate} \\
$g$    & 9.8 & m s $^{-2}$ & \\
$\rho_i$    & 900 & kg m$^{-3}$ & \\ %917
$ \rho_w $   & 1000 & kg m$^{-3}$ & \\
$C_W$    & $ 7.6\times 10^{6}$ & Pa m$^{-1/3}$ s$^{1/3}$  &   \\
$m_W$ & 1\slash3 & & \\ 
$A$ & $1 \times 10^{-25}$ ($6.8 \times 10^{-24}$--$3.6 \times 10^{-27}$) 
& s$^{-1}$ Pa$^{-3}$  & \citet{cuffey2010physics}  \\
$n$ & 3 & & \\ 
$C_C$ & 0.42 &  & \\ %\citep{tulaczyk2000basal}  \\
$E$ & $1 \times 10^{5}$  & Pa m  \\
$[m]$ & 1 & mm yr$^{-1}$ &   \\
$k$ & $4.0 \times 10^{-9}$ & m$^2$  \\
$\eta$ & $10^{-3}$  & Pa s \\
$k_{sb}$ & ($1 \times 10^{-17}$--$1 \times 10^{-12}$) & m$^2$ & \citet{freeze1979groundwater} \\
$\xi$ & 0.8 (0.3--1.0) &  & \citet{neuzil2014ice} \\
$S_s$ & ($1\times 10^{-7}$--$6.8\times10^{-4}$) & m$^{-1}$ & \citet{stone1997inversion, neuzil2014ice}\\
%$S_s$ & $1 \times 10^{-4}$ & m$^{-1}$ & \citet{robel2023contemporary}\\
% Range 1e-7 - 1e-4
\end{tabular}
\caption{Imposed parameters and scalings (given to 2 significant figures where applicable).}
\label{tab:paramsgiven}
\end{table}

\begin{table}
\centering
\begin{tabular}{ c| c| c}
Parameter     & Value Used & Units \\
\hline
$[H] $   & 2300 & m \\ 
$[t]$    & 7800 & yr \\
$[u]$    &  64 & m yr$^{-1}$ \\
$[N]$   &  $1.15 \times 10^{6}$  & Pa \\ % That's small
$[h]$   & 0.087 & m \\ 

\end{tabular}
\begin{tabular}{ c| c}
Parameter     & Value Used (Range)  \\
\hline

$ \delta $   & 0.1   \\
$\epsilon$    & $8.3\times 10^{-4}$   \\
$\lambda$    & $0.011$  \\
$\mathcal{E}$    & 0.05  \\ 
$\mathcal{C}$    & 5  \\
$K$    & ($6.1 \times 10^{-5}$--6)  \\
%$\Sigma$    & 5.1 \\
$\Sigma$    & ($0.07$--$480$) \\
\end{tabular}
\caption{Derived parameters and scalings (given to 2 significant figures where applicable).}
\label{tab:paramsderived}
\end{table}

In this section, we detail the procedure of non-dimensionalising the model presented in this paper. 
As we proceed, we will take certain scalings and parameters as imposed, and the values we choose are summarised in Table \ref{tab:paramsgiven}.
We will use these in turn to derive further scales and parameters, whose values are shown in Table \ref{tab:paramsderived}. 
%We begin with the case of the Weertman sliding law, in which the ice sheet is decoupled from the hydrological system, allowing us to consider the ice sheet and subglacial hydrology separately in turn. 
We begin with the ice sheet model, which consists of Equations \eqref{eq:massi} and \eqref{eq:momi}, combined with the sliding law \eqref{eq:rc}, and with the boundary conditions \eqref{eq:x0i}, \eqref{eq:Hixg} and \eqref{eq:stressxg}. 

We assume as in \citet{schoof2007marine}, that scales $[x]$ and $[a]$ for the horizontal length of the ice sheet and the accumulation rate are provided, the former due to (\textit{e.g.}) the size of the continental shelf or the regional extent of accumulation. 
We then obtain scales for the ice thickness $[H]$ and ice velocity $[u]$ by setting 
\begin{equation}
[H][u] = [a][x],  \quad C_W[u]^{m_W} = \rho_i g [H]^2 \slash [x].
\end{equation}
The first expression comes from the mass conservation equation \eqref{eq:massi}, and the second from assuming a balance between the basal and driving stress in the momentum equation \eqref{eq:momi}, with a Weertman or regularised Coulomb sliding law. 
The timescale is then given by
\begin{equation}
 [t]=[x]/[u]=[H]/[a].
\end{equation}

We may rescale the variables in Equations \eqref{eq:massi} and \eqref{eq:momi} using these scales, and scaling $b$ with $[H]$. 
We also assume we have a scale $[N]$ for the effective pressure, although we are presently yet to derive this. 
After this, we obtain the dimensionless equations
\begin{equation}\label{eq:massi*2}
  \frac{\partial {H}}{\partial t} +   \frac{\partial}{\partial x}{} (H u) = a,
\end{equation}
\begin{equation}\label{eq:momih*2}
4 \varepsilon   \frac{\partial}{\partial x}{} \left( H \left|  \frac{\partial{u}}{\partial x}\right|^{1/n-1}\frac{\partial{u}}{\partial x} \right) - \mathcal{C} N \left(\frac{ |u| }{|u| + (\mathcal{C} N )^{1/m_W} }\right)^{m_W} \text{sgn}(u))  - H   \frac{\partial}{\partial x}{}(H+b) = 0,
\end{equation}
where the dimensionless parameters are given by
\begin{equation}
\varepsilon = \frac{A^{-1/n} ([u]/[x])^{1/n} }{2 \rho_i g [H]} \ll 1.
\end{equation}
and
\begin{equation}
\mathcal{C} = \frac{C_C [N]}{C_W [u]^{m_W}}. 
\end{equation} 
%We shall see that $\mathcal{C}$ is $O(1)$ for our chosen values. 
In addition, the boundary conditions \eqref{eq:x0i}, \eqref{eq:Hixg} and \eqref{eq:stressxg} become
\begin{equation}\label{eq:x0i*}
u =0 \quad \text{at} \quad x=0.
\end{equation}
\begin{equation}\label{eq:Hixg*}
H = - \frac{b}{1-\delta} \quad \text{at} \quad x=x_g.
\end{equation}
\begin{equation}\label{eq:stressxg*}
4\varepsilon H \left|  \frac{\partial {u} }{\partial x}\right|^{1/n-1}   \frac{\partial {u} }{\partial x}= \frac12 \delta H^2  \quad \text{at} \quad x=x_g,
\end{equation}
where
\begin{equation}
\delta = 1- \frac{\rho_i}{\rho_w} \approx 0.1.
\end{equation}

We next seek to non-dimensionalise the subglacial hydrology equation \eqref{eq:hpre}. 
In addition to the scales described above, we assume that a scale for subglacial water sources $[m]$ is provided, \textit{e.g.} by considering the melt rate expected from estimated geothermal heat fluxes beneath Antarctica \citep{mccormack2022fine}. 
We scale $q_E$ with $[m]$, working on the assumption that meltwater fluxes and exfiltration contribute similar amounts to subglacial water budgets  \citep{christoffersen2014significant}. 
We also assume that, away from the grounding line, the effective pressure is relatively small compared to the overburden, an assumption consistent with borehole data from West Antarctica \citep{engelhardt1997basal}. %the results of other subglacial hydrology modelling \citep[\textit{e.g.}][]{ehrenfeucht2025antarctic}. 
%This means that the size of the hydraulic potential $\psi$ is dominated by the ice overburden, which scales with $\rho_i g [H]$. 
This means that we can choose a hydrostatic pressure scale $[p] = \rho_w g [H]$. 
Assuming that the source terms $m$ and $q_E$ are balanced by the flux divergence in Equation \eqref{eq:hpre} (as is required for a steady state), we can then derive a scale for the subglacial water thickness $h$, and subsequently the effective pressure $N$ via Equation \eqref{eq:Nh}
\begin{equation}\label{eq:scalehN}
%[h]= \frac{\eta [x]^2 [m] }{k \rho_i g [H]}, \quad [N] = \frac{E}{[h]}.
[h]= \frac{\eta [x]^2 [m] }{k \rho_w g [H]}, \quad [N] = \frac{E}{[h]}.
\end{equation}
For a general relation $h(N)$, $[N]$ can be determined by inverting this relation to find the $[N]$ such that $[h] = h([N])$. 
Under these scalings, the mass conservation equation \eqref{eq:hpre} and closure relation \eqref{eq:Nh} become
\begin{equation}\label{eq:hyd*}
%\lambda   \frac{\partial h}{\partial t} =   \frac{\partial}{\partial x}{} \left[ h   \frac{\partial}{\partial x}{} \left( H - \mathcal{E}N + \frac{b}{1-\delta}\right)  \right] + m + q_E, \quad N = \frac{1}{h},
\lambda   \frac{\partial h}{\partial t} =   \frac{\partial}{\partial x}{} \left[ h   \frac{\partial}{\partial x}{} \left( (1-\delta) H - \mathcal{E}N + b\right)  \right] + m + q_E, \quad N = \frac{1}{h},
\end{equation}
where the dimensionless parameters are given by
\begin{equation}
\lambda = \frac{[h]}{[m][t]} = \frac{[a][h]}{[m][H]} \ll 1,
\end{equation}
and
\begin{equation}
%\mathcal{E} = \frac{[N]}{\rho_i g [H]}. % * why is this called E now?
\mathcal{E} = \frac{[N]}{\rho_w g [H]}.
\end{equation}
The boundary conditions \eqref{eq:hx0} and \eqref{eq:Nxg} become: 
\begin{equation}\label{eq:hx0*}
%h   \frac{\partial}{\partial x}{} \left(H - \mathcal{E}N + \frac{b}{1-\delta}\right) = 0 \quad \text{at} \quad x=0,
h   \frac{\partial}{\partial x}{} \left((1-\delta) H - \mathcal{E}N + b\right) = 0 \quad \text{at} \quad x=0,
\end{equation}
\begin{equation}\label{eq:Nxg*}
N = 0 \quad \text{at} \quad x=x_g.
\end{equation}
The fact that $\lambda$ is small (see Table \ref{tab:paramsderived}) suggests that our assumption of a balance between the flux divergence and source terms in Equation \eqref{eq:hpre} is appropriate. 

% The constant $\mathcal{E}$ represents the characteristic size of the effective pressure relative to that of the ice overburden, which we expect to be fairly small based on existing modelled effective pressure beneath Antarctica (\textit{e.g.} \citet{ehrenfeucht2025antarctic}). 
% We choose $\mathcal{E}=0.05$, based on the the previous modelling assumption by \citet{bueler2009shallow} that subglacial effective pressure is at most 5\% of the overburden. 
% * We don't consider varying E because it's equivalent to varying other parameters?

% * Notice that we've repeated a sentence?
In considering the exfiltration, we scale $H_{sb}$ using the ice sheet thickness scale $[H]$, since sedimentary basins are typically hundreds to thousands of metres deep \citep{gustafson2022dynamic, tankersley2022basement}, and otherwise use the scalings provided in Tables \ref{tab:paramsgiven} and \ref{tab:paramsderived}. 
The dimensionless version of Equation \eqref{eq:qE} for $q_E$ is then
\begin{equation}\label{eq:qE*}
%q_E =  K   \frac{\partial}{\partial x}{} \left[  H_{sb}   \frac{\partial}{\partial x}{} \left( H - \mathcal{E} N +  \frac{b}{1-\delta} \right) \right] - \Sigma H_{sb}   \frac{\partial}{\partial t}{}\left(H - \mathcal{E} N\right),
q_E =  K   \frac{\partial}{\partial x}{} \left[  H_{sb}   \frac{\partial}{\partial x}{} \left( H - \mathcal{E} N +  \frac{b}{1-\delta} \right) \right] - \Sigma H_{sb}   \frac{\partial}{\partial t}{}\left( (1-\delta)(1-\xi)H - \mathcal{E} N\right),
\end{equation} 
where the dimensionless parameters are given by
\begin{equation}
%K = \frac{k_{sb} [H]}{k [h]} = \frac{k_{sb} [H]^2 \rho_i g}{\eta [x]^2 [m] } ,
K = \frac{k_{sb} [H]}{k [h]} = \frac{k_{sb} [H]^2 \rho_w g}{\eta [x]^2 [m] } ,
\end{equation}
and
\begin{equation}
%\Sigma = \frac{\rho_i S_s (1-\xi) [H]^2}{\rho_w [m][t]}. 
\Sigma = \frac{ S_s  [H]^2}{ [m][t]}. 
\end{equation} 
We note in particular that $K$ is independent of the cavity sheet permeability $k$. This is because the flux through the cavity sheet is determined by the melt rate, so that $k$ only influences $[h]$ and by extension $\lambda$ and $\mathcal{E}$, the latter of which can be adjusted by changing $E$.  
We note that not all parameters in this model are independent. In particular, due to Equation \eqref{eq:Nh}, rescaling $\mathcal{E}$ in Equation \eqref{eq:hyd*} is equivalent to simultaneously rescaling $m$, $K$ and $\Sigma$, up to a scaling of $h$.

\section{Non-Dupuit limit}

The use of the Dupuit approximation requires that both the aspect ratio $\alpha = [H]/[x] \ll 1$ and $S_s \eta [H]^2 / k_{sb} \rho_w g[t] \ll 1$. 
The latter fails if the hydraulic diffusivity is small, or if the timescales under consideration are short. 
In terms of the dimensionless variables described above, this failure occurs if $\Sigma / K$ is large ($O(\alpha^{-2})$). 
Although it is not the main focus of our work, we briefly consider what happens to the model in this case. 

We first note that, after non-dimensionalising using the scalings described in Appendix A, 
Equations \eqref{eq:groundwater} and \eqref{eq:qE0} become
\begin{equation}\label{eq:p*}
\Sigma\frac{\partial p_{sb}}{\partial t}  = K\left(\frac{\partial^2 p_{sb}}{\partial x^2} + \frac1{\alpha^2}\frac{\partial^2 p_{sb}}{\partial z^2} \right) + \Sigma\xi (1-\delta)  \frac{\partial H}{\partial t},
\end{equation}
\begin{equation}\label{eq:qE0*}
q_E =  -\frac{K}{\alpha^2} \left(\frac{\partial p_{sb}}{\partial z} + 1 - \alpha^2 \frac{\partial p_{sb}}{\partial x} \frac{\partial b}{\partial x}  \right) \bigg|_{z=b},
\end{equation}
solved subject to the boundary conditions (Equation \eqref{eq:bottomflux})
\begin{equation}\label{eq:pBC*}
p = p_w \quad \text{on} \quad z=b, \quad \frac{\partial p_{sb}}{\partial z} + 1 - {\alpha^2}\frac{\partial p_{sb}}{\partial x}\left(\frac{\partial b}{\partial x} - \frac{\partial H_{sb}}{\partial x} \right)  = 0 \quad \text{on} \quad z=b-H_{sb}.
\end{equation}
If $\Sigma \alpha^2/ K \ll1$, as we have assumed in order to use the Dupuit approximation, Equation \eqref{eq:p*} implies that $\partial^2 p_{sb} /\partial z^2 = O(\alpha^2)$, reflecting the fact that the pressure is hydrostatic at leading order under this approximation. 

To investigate the non-Dupuit case, we use a rescaled time variable $\tau = K t / \Sigma \alpha^2$. 
Substituting this timescale into Equation \eqref{eq:p*} gives
\begin{equation}\label{eq:pfast}
\frac{\partial p_{sb}}{\partial \tau}  = \alpha^2\frac{\partial^2 p_{sb}}{\partial x^2} + \frac{\partial^2 p_{sb}}{\partial z^2} + \xi (1-\delta)  \frac{\partial H}{\partial \tau},
\end{equation}
while Equations \eqref{eq:qE0*} and \eqref{eq:pBC*} are unchanged. 
Under this rescaling, the horizontal flux becomes negligible due to the prefactor $\alpha^2$, and we recover the model used by \citet{robel2023contemporary}, where the pressure satisfies a forced heat equation in one dimension.  
In this case, the vertical groundwater flow undergoes a time-dependent response to loading changes. \citet{robel2023contemporary} obtain analytical solutions for $p_{sb}$ and $q_E$ for the cases of a step change in $H$ and for a constant thinning / thickening rate $\partial H/\partial \tau$. 

In general, the solution of Equation \eqref{eq:pfast}, after neglecting the small horizontal fluxes, can be analysed using a Laplace transform in time. 
We use $s$ to denote the transform variable and overbars to denote the Laplace transform. 
Assuming that at time $\tau=0$ the basal pressure and ice thickness are $p_{w0}(x)$ and $H_{0}(x)$, and that the pressure is initially hydrostatic, Equation \eqref{eq:pfast} becomes
\begin{equation}\label{eq:pfastL}
s \bar{p}_{sb} - (p_{w0} + b-z) =  \frac{\partial^2 \bar{p}_{sb}}{\partial z^2} + \xi (1-\delta)  ( s\bar{H} - H_{0}),
\end{equation}
and Equation \eqref{eq:qE0*} becomes
\begin{equation}
\bar{q}_E =  -\frac{K}{\alpha^2} \left. \left(\frac{\partial{\bar{p}_{sb}}}{\partial z} + \frac{1}{s}  \right) \right|_{z=b}.
\end{equation}
The boundary conditions are
\begin{equation}
\bar{p}_{sb} =\bar{p}_w \quad \text{on} \quad z=b, \quad 
\frac{\partial{ \bar{p}_{sb}}}{\partial z} + \frac1s =0 \quad \text{on} \quad z=b-H_{sb}.
\end{equation}
Equation \eqref{eq:pfastL} can be solved to give
\begin{equation}
 \bar{p}_{sb} = \left( \bar{p}_w - \xi(1-\delta) \left(\bar{H} + \frac{H_{0}}{s}\right) \right)\frac{\cosh((z-b+H_{sb})\sqrt{s})}{\cosh(H_{sb}\sqrt{s})}  + \frac{b-z}{s} + \xi(1-\delta)  \left( \bar{H} - \frac{H_{0}}{s}\right),
\end{equation}
resulting in the transformed exfiltration
\begin{equation}
\bar{q}_E =  -\frac{K}{\alpha^2}  \left( \bar{p}_w + \frac{p_{w0}}{s} - \xi(1-\delta)\left(\bar{H} + \frac{H_{0}}{s}\right) \right)\sqrt{s} \tanh(H_{sb}\sqrt{s}).
\end{equation}
The solution for $p_{sb}$ and $q_E$ is then given by inversion of the Laplace transform. It can be shown using the final value theorem for the Laplace transform that, in the limit $\tau \to \infty$, this solution tends to the compaction-driven exfiltration found in the Dupuit solution \eqref{eq:qE*} above.  
Namely, we have that
\begin{align}
\lim_{\tau \to \infty} q_E &= \lim_{s\to0} s \bar{q}_E \\&= -\frac{K}{\alpha^2} H_{sb} \lim_{s\to0} \left( s\bar{p}_w + p_{w0} - \xi(1-\delta)\left(s\bar{H} + H_{0}\right) \right).
\\&= -\frac{K}{\alpha^2} H_{sb} \lim_{\tau \to \infty}\left( \frac{\partial p_w}{\partial \tau} - \xi (1-\delta)\frac{\partial H}{\partial \tau} \right)  \\&= \Sigma H_{sb} \left( \frac{\partial p_w}{\partial t} - \xi (1-\delta)\frac{\partial H}{\partial t} \right),
\end{align}
using the fact that $\tau = K t / \Sigma\alpha^2 $. The Dupuit approximation and short-time solution are therefore consistent limiting approximations of the groundwater flow problem. 

\renewcommand{\thefigure}{A\arabic{figure}}
\setcounter{figure}{0}

% \begin{figure}[t]
%     \centering
%     \includegraphics[width=0.75\linewidth]{figures/qg_ret_adv.png}
%     \caption{Flux across the grounding line $q_g =  H(x_g,t)u(x_g,t)$ plotted against grounding line position $x_g$, for (a) the retreat scenario in Figure \ref{fig:retreat} (b) the advance scenario  in Figure \ref{fig:advance}, for the same values of $S_s$.}
%     \label{fig:qg_ret_adv}
% \end{figure}

% In this section we include some additional figures, which support our discussion in Section 5 of the effect of compaction-driven exfiltration during grounding line motion. 
% That section focused on the grounding line retreat following a step decrease in the accumulation $a$, whose results were shown in Figure \ref{fig:retreat}. 

\section{Numerical method}

In this appendix, we describe the numerical method used to obtain the solutions above.
We discretise the equations spatially using finite volume schemes, and evolve them forward in time using a one-step fully implicit method. 
Numerical methods are implemented in MATLAB, with the inbuilt `fsolve' function used to solve the implicit equations at each time step. 

Since these equations are solved on the variable domain $0<x<x_g(t)$, we use the coordinate transformation 
\begin{equation}\label{eq:xtransform}
x = x_g(t) X,
\end{equation}   
which changes the domain to $0<X<1$, while introducing $x_g$ and $\dot{x}_g = \mathrm{d}x_g /\mathrm{d}t$ into the equations themselves. 

\begin{figure}[t]
    \centering
    \includegraphics[width=0.6\linewidth]{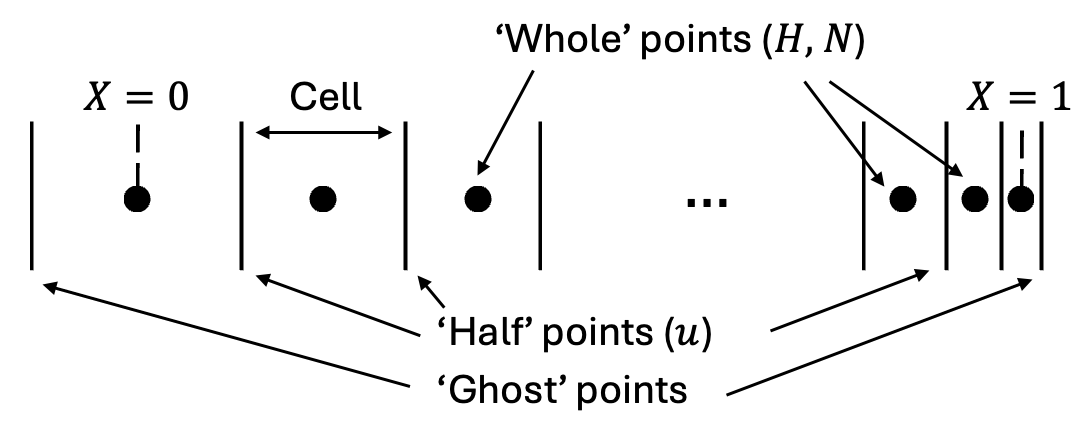}
    \caption{Schematic of the staggered mesh grid used to obtain numerical solutions for $H$, $u$, $x_g$ and $N$.}
    \label{fig:numerics_schematic}
\end{figure}

The shallow-shelf ice sheet model is governed by Equations \eqref{eq:massi*} and \eqref{eq:momih*}. Our procedure for solving these equations numerically is based on that described in \cite{schoof2007ice}. After applying the transformation \eqref{eq:xtransform}, Equation \eqref{eq:massi} becomes
\begin{equation}\label{eq:massi_disc}
   \frac{\partial H}{\partial t} - \frac{\dot{x}_g}{x_g} X  \frac{\partial H}{\partial X} +   \frac{1}{x_g}\frac{\partial}{\partial X}{} (Hu) = a.
\end{equation}
% where we define the operator
% \begin{equation}
%    \frac{D}{D t} = \frac{\partial}{\partial t} - \frac{\dot{x}_g}{x_g} X  \frac{\partial  }{\partial X}.
% \end{equation}
Equation \eqref{eq:massi_disc} is discretised using a first-order upwind scheme. 
The velocity $u$ is calculated at `half' mesh points, which are spaced halfway between the `whole' mesh points at which $H$ is found. 
Figure \ref{fig:numerics_schematic} shows a depiction of the staggered mesh grid used for this numerical method.  
We solve Equation \eqref{eq:massi_disc} at every `whole' mesh point, and in addition impose that $H = b /(1-\delta)$ at the rightmost mesh point $X=1$ (Equation \eqref{eq:Hixg*}), which implicitly determines the evolution of $x_g$. 

The ice velocity $u$ is found by solving the transformed version of Equation \eqref{eq:momih*} at each time step,
\begin{equation}
\label{eq:momi_disc}
\frac{4 \varepsilon}{{x_g}^{1/n+1}}   \frac{\partial}{\partial X}{} \left( H\left(  \left(\frac{\partial{u}}{\partial X } \right)^2 + \varepsilon_R^2\right)^{(1-n)/2n} \frac{\partial{u}}{\partial X} \right) - \mathcal{C} N \left(\frac{ |u| }{|u| + (\mathcal{C} N )^{1/m_W} }\right)^{m_W} \text{sgn}(u)  - \frac{1}{x_g} H \frac{\partial}{\partial X}{}(H+b) = 0.
\end{equation}
Following \citet{haseloff2025subglacial}, we introduce a regularisation parameter $\varepsilon_R=0.01$ to ensure smoothness when $\partial u / \partial X=0$. 
We impose discretised versions of the boundary conditions \eqref{eq:x0i*} and \eqref{eq:stressxg*} at `ghost' mesh points just outside the domain on either side. 
In discretising Equation \eqref{eq:momi_disc}, we estimate spatial derivatives $\partial/ \partial X$ using a central difference. 
The factor of $H$ in the driving stress is found as a midpoint arithmetic mean, as is $N$ in the basal friction. 

The evolution of the subglacial hydrological system is given by Equation \eqref{eq:hyd*2}. Applying the transformation \eqref{eq:xtransform}, this becomes
\begin{multline}\label{eq:hyd_disc}
-\left(\frac{\lambda}{N^2}    + \Sigma\mathcal{E} H_{sb} \right) \left(\frac{\partial N}{\partial t} - \frac{\dot{x}_g}{x_g} X  \frac{\partial  N}{\partial X} \right)=   \frac1{{x_g}^2}\frac{\partial}{\partial X}{} \left( \left(\frac1{N} + K H_{sb}\right)    \frac{\partial}{\partial X}{} \left( (1-\delta)  H- \mathcal{E} N +  b\right)  \right) + m \\ - \Sigma H_{sb}  (1-\delta)(1-\xi)\left(\frac{\partial H}{\partial t} - \frac{\dot{x}_g}{x_g} X  \frac{\partial  H}{\partial X} \right) .
\end{multline}
We implement this equation via a finite volume method, with $N$ solved at `whole' meshpoints, using the same grid as for $H$. 
Equation \eqref{eq:hyd_disc} is solved at all except the rightmost mesh point, where $N=0$ is imposed (Equation \eqref{eq:Nxg*}). 
In calculating the fluxes between cells, $1/N$ is estimated using its left-hand value in order to avoid a singularity from the rightmost grid cell. The flux into the leftmost cell is imposed to be minus the outgoing flux, to satisfy the zero-flux condition at $x=0$ (Equation \eqref{eq:hx0*}).

We solve Equations \eqref{eq:massi_disc},  \eqref{eq:momi_disc} and \eqref{eq:hyd_disc} on a 300-point variable grid, with mesh points more tightly spaced towards the grounding line and a mesh spacing decreasing to $10^{-4}$ in the rightmost cell. 
This ensures that the extensional stress boundary layer near the grounding line is accurately resolved. 
The numerical method for the ice sheet was validated for the case of a Weertman sliding law using the results of \cite{schoof2007ice}, and the method for the hydrology was validated against the exact solution for $K=0$ given by Equation \eqref{eq:Nexact}.

Because steady states of the shallow-shelf model are non-unique in general, we obtain steady-state solutions by evolving the model forward in time until successive time steps satisfy (in discretised form)
\begin{equation}
\int_0^{x_g} \left( \left( \frac{\partial H}{\partial t}\right)^2 + \left( \frac{\partial u}{\partial t}\right)^2 + \left( \frac{\partial N}{\partial t}\right)^2 \right)\, \mathrm{d}x + \left( \frac{\mathrm{d} x_g}{\mathrm{d} t}\right)^2 < 0.001^2.
\end{equation}
In order to promote fast convergence, we use an artificially small value of the parameter $\lambda$ when finding steady-state solutions.  
To obtain an initial condition, we solve the approximate steady state equation for $x_g$ found by \citet{tsai2015marine}, and integrate the ice sheet model while ignoring the extensional stress terms. 
To obtain the time-dependent solutions in Section 5, we use variable time stepping, with shorter time steps early on in order to accurately resolve the initial behaviour.

\end{document}